\documentclass[submission,copyright]{eptcs}
\providecommand{\event}{QPL 2015} % Name of the event you are submitting to
\usepackage[T1]{fontenc}
\usepackage{textcomp}
\usepackage[utf8]{inputenc}

\usepackage{graphicx}
\usepackage{latexsym}
\usepackage{amssymb,amsmath}
\usepackage{mathtools}
\usepackage[amsmath,thmmarks,hyperref]{ntheorem}
\usepackage{stmaryrd}
\usepackage[noadjust]{cite}
\usepackage{bussproofs}
\usepackage{enumitem}
\usepackage{array}
\usepackage{color}

\usepackage{comment} % for auxproof
\usepackage{wrapfig}
\usepackage{picins}

\allowdisplaybreaks[1] % amsmath option: allows page breaks inside a
\usepackage[all,2cell]{xy}
\UseAllTwocells
\SelectTips{cm}{}  % Tips (of arrows) are in accordance with Computer
\newdir{pb}{:(1,-1)@^{|-}}
  \def\pb#1{\save[]+<17 pt,0 pt>:a(#1)\ar@{pb{}}[]\restore}
\input{mysetting-ntheorem}

\newcommand\myqed{\qed}

\newif\ifignore
\ignorefalse % no auxproof
\let\emptyset\varnothing

\DeclareMathSymbol{\mhyphen}{\mathalpha}{operators}{"2D}
\ifdefined\mathsl\else
  \DeclareMathAlphabet{\mathsl}{\encodingdefault}{\rmdefault}{\mddefault}{\sldefault}
  \SetMathAlphabet{\mathsl}{bold}{\encodingdefault}{\rmdefault}{\bfdefault}{\sldefault}
\fi
\DeclareFontFamily{U}{mathux}{\hyphenchar\font45}
\DeclareFontShape{U}{mathux}{m}{n}{
      <5> <6> <7> <8> <9> <10>
      <10.95> <12> <14.4> <17.28> <20.74> <24.88>
      mathux10
      }{}
\DeclareSymbolFont{mathux}{U}{mathux}{m}{n}

\DeclareMathSymbol{\bigovee}{1}{mathux}{"8F}
\DeclareMathSymbol{\bigperp}{1}{mathux}{"4E}

\newcommand{\To}{\Rightarrow}

\newcommand{\longto}{\longrightarrow}
\newcommand{\longgets}{\longleftarrow}

\newcommand{\C}{\mathbb{C}}

\newcommand{\Set}{\mathbf{Set}}

\newcommand{\GEMod}{\mathbf{GEMod}}
\newcommand{\Pfn}{\mathbf{Pfn}}

\newcommand{\Conv}{\mathbf{Conv}}
\newcommand{\SConv}{\mathbf{SConv}}

\newcommand{\PCM}{\mathbf{PCM}}

\newcommand{\Cstar}{\mathbf{Cstar}}

\newcommand{\EMod}{\mathbf{EMod}}

\newcommand\Eff{\mathbf{Eff}}
\newcommand\FPE{\mathbf{FPE}}

\newcommand{\PU}{\mathrm{PU}}
\newcommand{\PSU}{\mathrm{PSU}}

\newcommand{\Kl}{\mathcal{K}\mspace{-2mu}\ell}
\newcommand{\EM}{\mathcal{E}\mspace{-3mu}\mathcal{M}}
\newcommand{\id}{\mathrm{id}}

\newcommand{\Pow}{\mathcal{P}}

\newcommand{\Dst}{\mathcal{D}}

\makeatletter
\newcommand*\mywidehat[1]{\mathpalette\mywidehathelper{#1}}
\newcommand*\mywidehathelper[2]{%
        \hbox{\dimen@\accentfontxheight#1%
                \accentfontxheight#11.2\dimen@
                $\m@th#1\skew{1}\widehat{#2}$%
                \accentfontxheight#1\dimen@
        }%
}
\newcommand*\accentfontxheight[1]{%
        \fontdimen5\ifx#1\displaystyle
                \textfont
        \else\ifx#1\textstyle
                \textfont
        \else\ifx#1\scriptstyle
                \scriptfont
        \else
                \scriptscriptfont
        \fi\fi\fi3
}
\makeatother

\newcommand{\Dsub}{\mywidehat{\mathcal{D}}}

\newcommand{\op}{\mathrm{op}}
\newcommand{\bang}{\mathord{!}}

\DeclarePairedDelimiter\ket{\lvert}{\rangle}
\DeclarePairedDelimiter\cotup{[}{]}

\DeclarePairedDelimiter\nset{[}{]}
\newcommand{\lft}{\mathopen{}\mathclose\bgroup\left}
\newcommand{\rgt}{\aftergroup\egroup\right}
\newcommand{\mdl}{\mathrel{}\aftergroup\mathrel\aftergroup{\aftergroup}\middle}

\newcommand{\cat}[1]{\mathbf{#1}}
\newcommand{\catA}{\cat{A}}
\newcommand{\catB}{\cat{B}}
\newcommand{\catC}{\cat{C}}
\newcommand{\catD}{\cat{D}}
\newcommand{\catE}{\cat{E}}

\DeclareMathSymbol{\hatsym}{\mathalpha}{operators}{"5E}
\newcommand{\Laccent}{\hat}

\newcommand{\Lcirc}{\mathbin{\Laccent{\circ}}}
\newcommand{\Lplus}{\mathbin{\Laccent{+}}}
\newcommand{\Lid}{\Lin{\id}}
\newcommand{\Lcopr}{\Lin{\kappa}}

\newcommand{\Lto}{\rightharpoonup}

\newcommand{\Lin}[1]{\widehat{#1}}
\newcommand{\Linsym}{\widehat{\ }}

\newcommand\KlL[1]{{#1_{+1}}}
\newcommand\catBL{\KlL{\catB}}

\newcommand\totcat[1]{{#1_\mathrm{t}}}

\DeclareMathOperator{\Dp}{Dp}

\newcommand{\Pred}{\mathsl{Pred}}
\newcommand{\Stat}{\mathsl{Stat}}
\newcommand{\SStat}{\mathsl{SStat}}

\newcommand{\pproj}{\mathord{\vartriangleright}}

\newcommand{\smul}{\mathbin{\bullet}}

\newcommand{\mathinvsymbolsub}[2]{\rotatebox[origin=c]{180}{$#1#2$}}
\newcommand{\mathinvsymbol}[1]{\mathpalette\mathinvsymbolsub{#1}}
\newcommand{\invbang}{\mathord{\mathinvsymbol{!}}}

\newcommand{\Iff}{\Longleftrightarrow}

\newcommand\instr{\mathsf{instr}}

\newcommand\makexypartial{\everyentry={\let\tmpar\ar
\renewcommand\ar{\tmpar@{-^>}}}}

\DeclareMathOperator{\dom}{dom}

\newcommand{\auxproof}[1]{%
\ifignore\mbox{}\newline
\textbf{BEGIN: AUX-PROOF}\dotfill\newline
{\footnotesize#1}\mbox{}\newline
\textbf{END: AUX-PROOF}\dotfill\newline
\fi}

\ifignore

\else
\excludecomment{auxproofenv}
\fi

\newcommand{\urlalt}[2]{\href{#1}{\urlstyle{rm}\nolinkurl{#2}}}

\DeclareMathAlphabet{\mathcal}{OMS}{cmsy}{m}{n}
\SetMathAlphabet{\mathcal}{bold}{OMS}{cmsy}{b}{n}

\DeclareSymbolFont{cmlargesymbols}{OMX}{cmex}{m}{n}
\DeclareMathSymbol{\cmcoprod}{\mathop}{cmlargesymbols}{"60}
\let\coprod\cmcoprod

\title{Total and Partial Computation\\
in Categorical Quantum Foundations}
\author{Kenta Cho
\institute{Institute for Computing and Information Sciences\\
Radboud University, Nijmegen, The Netherlands}
\email{%
\href{mailto:K.Cho@cs.ru.nl}{\nolinkurl{K.Cho@cs.ru.nl}}%
\textrm{,\enspace}%
\url{http://www.cs.ru.nl/K.Cho/}}
}
\def\titlerunning{Total and Partial Computation
in Categorical Quantum Foundations}
\def\authorrunning{K.~Cho}

\begin{document}

\maketitle

\begin{abstract}
This paper uncovers the fundamental relationship
between total and partial computation in the form of an equivalence
of certain categories. This equivalence involves on the one hand \emph{effectuses},
which are categories for \emph{total} computation,
introduced by Jacobs for the study of quantum/effect logic.
On the other hand, it involves what we call \emph{FinPACs with effects};
they are finitely partially additive categories equipped with effect algebra
structures, serving as categories for \emph{partial} computation.
It turns out that the Kleisli category of the lift monad $(-)+1$
on an effectus is always a FinPAC with effects, and this construction
gives rise to the equivalence.
Additionally, \emph{state-and-effect triangles} over FinPACs with effects
are presented.
\end{abstract}

% Typeset with \makeatletter
% \csname ver@xy.sty\endcsname
% \makeatother

%=====================
\section{Introduction}
\label{sec:intro}
%=====================

An \emph{effectus} is a category with a final object $1$
and finite coproducts $(+,0)$ satisfying certain assumptions
(see Definition~\ref{def:effectus}),
introduced recently by Jacobs~\cite{Jacobs2015NewDir},
which provides a suitable setting for quantum/effect logic and computation.
In an effectus, arrows $\omega\colon 1\to X$ are \emph{states}
on $X$, and $p\colon X\to 1+1$ are \emph{predicates}.
They turn out to form a \emph{convex set} and a so-called \emph{effect module},
respectively. Arrows $f\colon X\to Y$ are seen as computation,
inducing state and predicate transformers.
The situation is summarised in a \emph{state-and-effect triangle},
see \S{}\ref{subsec:effectus} for an overview of effectuses.

Motivating examples of effectuses,
which model quantum computation and logic,
are given by $C^*$-algebras with (completely)
positive unital maps, and by $W^*$-algebras with
normal (completely) positive unital maps.
Other effectuses include
the category $\Set$ of sets for a classical setting,
and the Kleisli category $\Kl(\Dst)$
of the distribution monad $\Dst$ for a probabilistic setting.
As seen in these examples,
computation modelled by an effectus is
\emph{total} (or terminating) but
not \emph{partial} (or non-terminating).
Indeed, arrows in an effectus always induce
`terminating' predicate transformers in
the sense that they preserve the truth predicates.
We need models of partial computation in some cases,
however, since programs do not necessarily terminate in general.
Moreover, such models often have richer structures such as
complete partial orders, which allow us to interpret loop and recursion.
For instance, the category of sets and \emph{partial} functions
are enriched over complete partial orders, and so is the category of $W^*$-algebras and
normal (completely) positive \emph{subunital} maps~\cite{Rennela2014,Cho2014QPL}.

The present paper studies partial computation in effectuses
via the \emph{lift monad} (a.k.a.\ maybe monad),
which is a common technique
in categorical semantics of computation, going back to Moggi~\cite{Moggi1991}.
We switch from an effectus $\catB$
to the Kleisli category of the lift monad $(-)+1$ on $\catB$,
which we denote by $\KlL{\catB}$.
An arrow $X\to Y$ in $\KlL{\catB}$ is
$X\to Y+1$ in $\catB$, seen
as a partial computation from $X$ to $Y$.
This simple idea makes a lot of sense for any effectus,
leading us to the main results of this paper as follows.
\begin{itemize}
\item
For an effectus $\catB$,
the Kleisli category $\KlL{\catB}$ of the lift monad
is a \emph{finitely partially additive category (FinPAC)},
which is a finite variant of Arbib and Manes' partially additive category
(PAC)~\cite{ArbibM1980,ManesA1986}.
The homsets $\KlL{\catB}(X,1)=\catB(X,1+1)$
are the sets of predicates and form effect algebras.
The category $\KlL{\catB}$ is what we call a \emph{FinPAC with effects},
which has an effect algebra structure related to the partially additive structure
in an appropriate manner (see Definition~\ref{def:finpac-with-effects}).
\item
On the other hand,
if $\catC$ is a FinPAC with effects,
then the subcategory $\totcat{\catC}$ with
`total' arrows is an effectus.
Moreover, the two constructions
$\KlL{(-)}$ and $\totcat{(-)}$
are inverses of each other up to isomorphism.
Categorically, we obtain a 2-equivalence
of the 2-categories of effectuses and FinPACs with effects.
\begin{equation}
\label{eq:equivalence-Eff-PEff}
\xymatrix{
\left(
\text{\begin{tabular}[c]{@{}c@{}}
effectuses \\
{}[\emph{total} computation]
\end{tabular}}
\right)
\ar@/^1.5ex/[rr]^{\KlL{(-)}}
&
\simeq
&
\ar@/^1.5ex/[ll]^{\totcat{(-)}}
\left(
\text{\begin{tabular}[c]{@{}c@{}}
FinPACs with effects \\
{}[\emph{partial} computation]
\end{tabular}}
\right)
}
\end{equation}
See Table~\ref{table:eff-P-eff}
for examples of
effectuses and corresponding FinPACs with effects.
\begin{table}[t]
\centering
\caption{Examples of effectuses and corresponding FinPACs with effects.}
\label{table:eff-P-eff}
\begin{tabular}{cccc}
effectus &
  total computation &
  partial computation &
  FinPAC with effects
\\ \hline\hline
$\catB$ &
  $X\to Y$ &
  $X\to Y+1$ (in $\catB$) &
  $\KlL{\catB}$
\\ \hline
$\Set$ &
  function &
  partial function &
  $\Pfn$
\\ \hline \noalign{\vskip .25ex}
$\Kl(\Dst)$ &
  \begin{tabular}[c]{@{}c@{}}
    $X\to \Dst Y$ \\[-.75ex]
    (stochastic relation)
  \end{tabular} &
  \begin{tabular}[c]{@{}c@{}}
    $X\to \Dsub Y$ \\[-.75ex]
    (substochastic relation)
  \end{tabular} &
  $\Kl(\Dsub)$
\\ \hline
$(\Cstar_\PU)^\op$ &
  positive unital map &
  positive subunital map &
  $(\Cstar_\PSU)^\op$
% \\ \hline
% $(\Wstar_\CPU)^\op$ &
%   \begin{tabular}[c]{@{}c@{}}
%     normal CP unital map, \\
%     or, quantum channel
%   \end{tabular} &
%   \begin{tabular}[c]{@{}c@{}}
%     normal CP subunital map, \\
%     or, quantum operation
%   \end{tabular} &
%   $(\Wstar_\CPSU)^\op$
\end{tabular}
\end{table}
This equivalence characterises
the Kleisli categories $\KlL{\catB}$ of the lift monad on
effectuses as FinPACs with effects, and
effectuses as the `total' subcategories $\totcat{\catC}$ of FinPACs with effects.
\end{itemize}

We additionally present two type of state-and-effect triangles
over a FinPAC with effects.
One triangle is rather simple and easy,
involving generalised effect modules and subconvex sets.
Another triangle is obtained by an application of
the above 2-equivalence to a state-and-effect triangle over an effectus,
but only under an additional `normalisation' condition.
This also contains a slight improvement of a known result
on effectuses with normalisation, via division in effect monoids.

The paper is organised as follows.
We first give preliminaries in the next section.
Section~\ref{sec:FinPAC} introduces FinPACs.
In \S\ref{sec:finpac-with-effects} we study partial computation in effectuses
and FinPACs with effects, and then in \S\ref{sec:equivalence-eff-FPE}
we prove categorical equivalence of effectuses and FinPACs with effects.
Section~\ref{sec:triangle-P-effectus} presents
state-and-effect triangles over FinPACs with effects.

%======================
\section{Preliminaries}
\label{sec:preliminaries}
%======================

%------------------------------------------
\subsection{Partial commutative monoids,
(generalised) effect modules and
(sub)convex sets}
%------------------------------------------

A \emph{partial commutative monoid (PCM)}
is a set $M$ with a partial binary `sum' operation
$\ovee\colon M\times M\rightharpoonup M$
and a `zero' element $0\in M$ subject to
$(x\ovee y)\ovee z\simeq x\ovee (y\ovee z)$,
$x\ovee y\simeq y\ovee x$ and $x\ovee 0\simeq x$,
where $\simeq$ denotes the Kleene equality:
if either side is defined, then so is the other, and they are equal.
We write $x\perp y$ if $x\ovee y$ is defined,
and we say elements $x_1,\dotsc,x_n$ are
\emph{orthogonal} if $x_1\ovee x_2\ovee\dotsb\ovee x_n$ is defined.
Any PCM carries a preorder via $x\le y
\,\Leftrightarrow\,
\exists z\ldotp x\ovee z=y$, with $0$ as a bottom
(a least element).
A \emph{generalised effect algebra (GEA)} is a PCM that is
positive ($x\ovee y=0\,\Rightarrow\, x=y=0$) and cancellative
($x\ovee y=x\ovee z\,\Rightarrow\, y=z$).
In a GEA the preorder $\le$ above is a partial order, and
we have a `partial difference' given by
$x\ominus y=z\,\Leftrightarrow\, x=y\ovee z$.
An \emph{effect algebra} is a GEA that has a top
(a greatest element), which is denoted by $1$.
Any element $x$ in an effect algebra has an orthocomplement
$x^\bot\coloneqq 1\ominus x$, i.e.\ a unique element
such that $x\ovee x^\bot=1$.
Homomorphisms of PCMs and GEAs preserve
$\ovee$ and $0$, and those of effect algebras
additionally preserve $1$.
An \emph{effect monoid} is an effect algebra $M$ with
a `multiplication' PCM-bihomomorphism ${\cdot}\colon M\times M\to M$
satisfying $1\cdot r=r=r\cdot 1$ and
$(r\cdot s)\cdot t=r\cdot (s\cdot t)$.
A \emph{partial commutative module (PCMod)}
over an effect monoid $M$ is a PCM $E$ with
a `scalar multiplication'
PCM-bihomomorphism ${\smul}\colon M\times E\to E$ satisfying
$1\smul x=x$ and $r\smul(s\smul x)=(r\cdot s)\smul x$.
A \emph{generalised effect module (GEMod)}
and an \emph{effect module}
are respectively
a GEA and an effect algebra that are at the same time a PCMod.
Homomorphisms of them are required to preserve the scalar multiplication.

For an effect monoid $M$,
we denote by $\Dst_M$ and $\Dsub_M$
respectively the (finite, discrete) distribution and
subdistribution monads over $M$ on the category $\Set$.
For a set $X$, the set $\Dst_M X$ consists of
formal convex sums $\ket{x_1}r_1+\dotsb+\ket{x_n}r_n$
where $x_i\in X$ and $r_i\in M$ with $\bigovee_i r_i= 1$,
while $\Dsub_M X$ consists of
$\ket{x_1}r_1+\dotsb+\ket{x_n}r_n$
with $\bigovee_i r_i\le 1$
(which holds automatically as long as
$r_1,\dotsc,r_n$ are orthogonal).
Here the `ket' notation $\ket{x}$ is
just syntactic sugar
to distinguish formal sums from elements $x\in X$.
A \emph{convex set} (resp.\ \emph{subconvex set}) over $M$
is an Eilenberg-Moore algebra of $\Dst_M$ (resp.\ $\Dsub_M$),
which is a set $X$ with an operation mapping
a formal sum $\sum_i \ket{x_i}r_i\in\Dst_M X$ (resp.\ $\Dsub_M X$)
to an actual sum $\bigovee_i x_ir_i\in X$.
For an effect monoid $M$
we write $\EMod_M$ and $\GEMod_M$ for the categories
of effect modules and GEMod's over $M$,
and $\Conv_M=\EM(\Dst_M)$ and $\SConv_M=\EM(\Dsub_M)$
for the categories of convex and subconvex sets over $M$.
The following dualities are fundamental.

\begin{myproposition}
\label{prop:emod-conv-duality}
There are the following adjunctions,
obtained by ``homming into $M$''.
\[
\xymatrix@C+1.5pc{
(\EMod_M)^\op \ar@/^1.5ex/[r]^-{\EMod(-,M)}
\ar@{}[r]|-{\top} &
\Conv_M \ar@/^1.5ex/[l]^-{\Conv(-,M)}
}\qquad
\xymatrix@C+1.5pc{
(\GEMod_M)^\op \ar@/^1.5ex/[r]^-{\GEMod(-,M)}
\ar@{}[r]|-{\top} &
\SConv_M \ar@/^1.5ex/[l]^-{\SConv(-,M)}
}
\]
\end{myproposition}
\begin{myproof}
The left-hand adjunction is shown in~\cite[Proposition~6]{Jacobs2015NewDir}.
The right-hand one is shown in \cite[Appendix~B]{Rennela2013Master}
for $M=[0,1]$, and the proof is easily generalised.
\end{myproof}

We say a PCMod $E$ over an effect monoid $M$ is \emph{subconvex} if
$r_1\smul x_1,\cdots,r_n\smul x_n$ are orthogonal
for any $x_1,\cdots,x_n\in E$ and for orthogonal $r_1,\cdots,r_n\in M$.
A subconvex PCMod is then
a subconvex set via the subconvex sum $\bigovee_i r_i\smul x_i$.
The category $\PCM$ of PCMs is symmetric monoidal closed
via a tensor product representing bihomomorphisms~\cite{JacobsM2012Coref}.
Therefore, $\PCM$-enriched categories are well-defined.
Explicitly, a category is $\PCM$-enriched
if each homset is a PCM and the composition is a
PCM-bihomomorphism.

%----------------------
\subsection{Effectuses}
\label{subsec:effectus}
%----------------------

Several assumptions on a category were identified by
Jacobs~\cite{Jacobs2015NewDir}
for the study of quantum/effect logic and computation.
A category that satisfies the most basic assumption~\cite[Assumption~1]{Jacobs2015NewDir}
is now called an `effectus', since~\cite{JacobsWW2015}.\footnote{%
Note that \cite[Definition~12]{JacobsWW2015} uses a slightly different
joint monicity requirement. In the present paper we follow~\cite{Jacobs2015NewDir}.}

\begin{mydefinition}
\label{def:effectus}
An \emph{effectus} is a category with
a final object $1$ and finite coproducts $(+,0)$ satisfying:
\begin{itemize}
\item
squares of the following form (E) and (K$^=$) are pullbacks;
\[
\xymatrix@R-1pc{
A+X\ar[r]^{\id+f} \ar[d]_{g+\id}
\ar@{}[dr]|{\textstyle\text{(E)}}
& A+Y \ar[d]^{g+\id} \\
B+X\ar[r]_{\id+f} & B+Y
}
\qquad\qquad
\xymatrix@R-1pc{
A\ar@{=}[r] \ar[d]_{\kappa_1}
\ar@{}[dr]|{\textstyle\text{(K$^=$)}} &
A \ar[d]^{\kappa_1} \\
A+X\ar[r]_{\id+f} & A+Y
}
\]
\item
the two arrows
$\cotup{\kappa_1,\kappa_2,\kappa_2},\cotup{\kappa_2,\kappa_1,\kappa_2}\colon
1+1+1\to 1+1$ are jointly monic.
\end{itemize}
\end{mydefinition}

In an effectus $\catB$,
a \emph{state} on an object $X$ is an arrow $\omega\colon 1\to X$;
a \emph{predicate} on $X$ is $p\colon X\to 1+1$;
and a \emph{scalar} is $r\colon 1\to 1+1$.
For a state $\omega$ and a predicate $p$,
the validity probability is given by
the \emph{abstract Born rule}
$(\omega\vDash p)\coloneqq p\circ\omega\colon 1\to 1+1$.
We write $\Stat(X)=\catB(1,X)$ and $\Pred(X)=\catB(X,1+1)$
for the sets of states and predicates respectively.

% \parpic[r]{%
% \begin{minipage}{.3\textwidth}
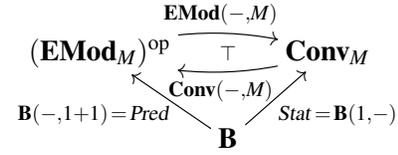
\begin{wrapfigure}{r}{.35\textwidth}
\vspace{-2.5ex}\centering
$
\xymatrix@C=1pc@R=1.5pc{
(\EMod_M)^\op \ar@/^1.5ex/[rr]^{\;\;\;\EMod(-,M)}
\ar@{}[rr]|-{\top}&
& \Conv_M \ar@/^1.5ex/[ll]^{\;\;\;\Conv(-,M)}
\\
&
\catB\ar[ul]^{\catB(-,1+1)\,=\,\Pred\;\;\;}
\ar[ur]_{\Stat\,=\,\catB(1,-)} &
}
$
\vspace{0ex}
\caption{State-and-effect triangle}
\label{fig:state-and-effect-triangle}
\end{wrapfigure}
% \end{minipage}}%
It turns out that the set of scalars $M\coloneqq \catB(1,1+1)$
is an effect monoid, and $\Stat(X)$ and $\Pred(X)$
are a convex set and an effect module over $M$ respectively.
In particular, $\Pred(X)$ is a poset with a top (truth) $1_X$ and
a bottom (falsum) $0_X$.
We refer to~\cite{Jacobs2015NewDir}
for the details, but later in~\S{}\ref{sec:triangle-P-effectus}
we will come to this point from a `partial' perspective.
An arrow $f\colon X\to Y$ in $\catB$
induces a state transformer $\Stat(f)\colon\Stat(X)\to\Stat(Y)$
by $\Stat(f)(\omega)=f\circ\omega$,
and a (backward) predicate transformer
$\Pred(f)\colon\Pred(Y)\to\Pred(X)$ by $\Pred(f)(p)=p\circ f$,
making $\Stat$ and $\Pred$ functors in a
\emph{state-and-effect triangle} shown in
Figure~\ref{fig:state-and-effect-triangle}.

The dual adjunction
$(\EMod_M)^\op\rightleftarrows\Conv_M$
from Proposition~\ref{prop:emod-conv-duality}
expresses the duality between predicates and states.
By ``currying'' the abstract Born rule
$\vDash\colon\Stat(X)\times\Pred(X)\to M$
we obtain maps $\alpha_X$ and $\beta_X$ in the
bijective correspondence of the dual adjunction:

\begin{prooftree}
\AxiomC{$\alpha_X\colon\Pred(X)\longto\Conv_M(\Stat(X),M)$ in $\EMod_M$}
\doubleLine
\UnaryInfC{$\beta_X\colon\Stat(X)\longto\EMod_M(\Pred(X),M)$ in $\Conv_M$}
\end{prooftree}

\noindent These maps $\alpha$ and $\beta$ are
natural transformations filling the triangle.

A motivating example of an effectus, which models quantum computation
and logic, is
the opposite $(\Cstar_\PU)^\op$ of the category of (unital) $C^*$-algebras
and positive unital (PU) maps.
Note that an initial object in $\Cstar_\PU$
is the set of complex numbers $\C$,
and finite products are given by the cartesian
products of underlying sets with coordinatewise operations;
they are a final object and finite coproducts
in the opposite. Then, states on a $C^*$-algebra $A$
are PU-maps $\omega\colon A\to\C$, which coincide
with the standard definition of `states' in operator theory.
Predicates on $A$ are PU-maps $f\colon \C\times\C\to A$,
which are in bijective correspondence with
elements $p\in A$ with $0\le p\le 1$,
via $p=f(1,0)$ and $f(\lambda,\rho)=\lambda p+\rho(1-p)$.
Such elements $p\in A$ with $0\le p\le 1$
are called \emph{effects} and thought of as
``unsharp'' predicates, which include
``sharp'' projections.
Scalars are effects in the complex numbers $\C$,
i.e.\ real numbers between $0$ and $1$. Then the abstract Born rule
is the usual Born rule $(\omega\vDash p)\coloneqq\omega(p)\in[0,1]$.
The sum $\ovee$ of effects $p,q$ is defined if $p+q\le 1$;
and in that case $p\ovee q=p+q$. With an obvious scalar multiplication,
effects form an effect module. The convex structure of states $\omega\colon A\to\C$
is given in a pointwise manner.
One has similar examples of effectuses given by
$C^*$-algebras with completely positive unital maps,
and $W^*$-algebras with normal (completely) positive unital maps.

Another example of an effectus is the category $\Set$
of sets and functions, which models classical computation and logic.
States $1\to X$ are simply elements $x\in X$,
while predicates $X\to 1+1\cong 2$ are subsets $P\subseteq X$ as usual.
The set of scalars is the two element set $\{0,1\}$,
and then the abstract Born rule is the membership
relation $(x\vDash P)=(x\in P)$.
The Kleisli category $\Kl(\Dst)$
of the distribution monad $\Dst=\Dst_{[0,1]}$ over $[0,1]$
is an effectus for a probabilistic setting.
States $1\to X$ are functions $1\to\Dst(X)$,
hence probability distributions $\omega\in\Dst(X)$.
Predicates $X\to 1+1$ are functions $X\to \Dst(1+1)\cong[0,1]$,
thus `fuzzy' predicates $p\in[0,1]^X$.
The set of scalars is the unit interval $[0,1]$,
and the abstract Born rule is given by the expectation value
$(\omega\vDash p)=\sum_x p(x)\omega(x)$.

Further explanation and examples are found in~\cite{Jacobs2015NewDir}.

%======================================================
\section{Finitely partially additive categories (FinPACs)}
\label{sec:FinPAC}
%======================================================

Here we introduce a notion of
\emph{finitely partially additive category (FinPAC)},
which is a finite variant of
Arbib and Manes' \emph{partially additive category (PAC)}~\cite{ArbibM1980,ManesA1986}.
The difference is that a PAC involves \emph{countable} sums, but
a FinPAC involves only \emph{finite} sums.

We first need a few preliminary definitions.
A category has \emph{zero arrows} if
there is a family of `zero arrows'
$0_{XY}\colon X\to Y$
such that $0_{WY}\circ f=0_{XW}
=g\circ 0_{XZ}$ for all $f\colon X\to Y$
and $g\colon Z\to W$.
Such a family is unique if exists
(indeed, $0_{XY}=0_{XY}\circ 0'_{XX}=0'_{XY}$).
If a category has a zero object $0$,
then it has zero arrows $X\to 0\to Y$.
The converse is also true
when the category has an initial (or final) object,
see e.g.~\cite[\S2.2.19]{ManesA1986}.
For a coproduct $\coprod_{i\in I} X_i$ in
a category with zero arrows,
we define \emph{partial projections}\footnote{%
Arbib and Manes call them \emph{quasi projections}.}
$\pproj_i\colon \coprod_{i\in I} X_i\to X_i$
by $\pproj_i\circ\kappa_i =\id_{X_i}$
and $\pproj_i\circ\kappa_j =0_{X_jX_i}$ ($j\ne i$).

\begin{mydefinition}[{cf.~\cite[\S3.3]{ArbibM1980}}]
\label{def:finpac}
A \emph{finitely partially additive category}
(\emph{FinPAC} for short)
is a category $\catC$ with finite coproducts $(+,0)$
which is $\PCM$-enriched
and satisfies the following two axioms.
\begin{itemize}
\item (Compatible sum axiom)
Arrows $f,g\colon X\to Y$
are orthogonal (in the PCM $\catC(X,Y)$)
whenever $f$ and $g$ are \emph{compatible}
in the sense that there exists a `bound'
$b\colon X\to Y+Y$ such that
$f=\pproj_1\circ b$ and $g=\pproj_2\circ b$.
\item (Untying axiom)
If $f,g\colon X\to Y$ are orthogonal,
then $\kappa_1\circ f,\kappa_2\circ g\colon X\to Y+Y$
are orthogonal too.
\end{itemize}
Note that $\catC$ has zero arrows,
i.e.\ zero elements $0_{XY}$ of the PCMs $\catC(X,Y)$.
\end{mydefinition}

For any objects $Y_1$ and $Y_2$ in a FinPAC,
arrows $\kappa_1\circ\pproj_1,\kappa_2\circ\pproj_2
\colon Y_1+Y_2\to Y_1+Y_2$ are compatible via a bound $\kappa_1+\kappa_2$,
hence orthogonal.
Then we obtain $\kappa_1\circ\pproj_1\ovee\kappa_2\circ\pproj_2=\id_{Y_1+Y_2}$
because $(\kappa_1\circ\pproj_1\ovee\kappa_2\circ\pproj_2)\circ\kappa_i=\kappa_i$.
For any arrow $f\colon X\to Y_1+Y_2$, therefore, one has a `decomposition'
$f=(\kappa_1\circ\pproj_1\ovee\kappa_2\circ\pproj_2)\circ f=
\kappa_1\circ f_1 \ovee \kappa_2\circ f_2$,
where $f_i=\pproj_i\circ f\colon X\to Y_i$.
It then easily follows that
the two partial projections $\pproj_i\colon Y_1+Y_2\to Y_i$ are jointly monic:
$\pproj_i\circ f=\pproj_i\circ g$ for $i=1,2$ implies $f=g$.
Now we see the $\PCM$-enrichment of a FinPAC is unique,
as is the case for a PAC (cf.~\cite[Theorem~3.2.18]{ManesA1986}).

\begin{myproposition}
\label{prop:unique-pcm-enrichment}
In a FinPAC,
arrows $f,g\colon X\to Y$ are orthogonal if and only if
they are compatible. In that case,
a bound $b\colon X\to Y+Y$ for $f$ and $g$ is unique and
$f\ovee g=\nabla\circ b$,
where $\nabla=[\id,\id]$ is the codiagonal.
\end{myproposition}
\begin{myproof}
Assume that $f$ and $g$ are orthogonal.
By untying, $\kappa_1\circ f, \kappa_2\circ g\colon X\to Y+Y$
are orthogonal, and then let $b=\kappa_1\circ f\ovee \kappa_2\circ g$.
It is easy to see that $f$ and $g$ are compatible via $b$.
A bound $b\colon X\to Y+Y$ is unique since $\pproj_1$
and $\pproj_2$ are jointly monic.
Finally, $\nabla\circ b=\nabla\circ
(\kappa_1\circ f\ovee \kappa_2\circ g)=f\ovee g$.
\end{myproof}

In a FinPAC, not very surprisingly, the $n$-ary version
of the compatible sum and the untying axiom hold, proved by induction
with a small trick; see Lemma~\ref{lem:n-ary-finpac-axioms}.
Therefore, we have the decomposition property for finite coproducts,
shown in the same way as the binary case above.

\begin{mylemma}
\label{lem:FinPAC-nary-decomp}
Let $\coprod_{i} Y_i$ be a finite coproduct
in a FinPAC. Any arrow $f\colon X\to\coprod_{i} Y_i$
is uniquely decomposed as $f=\bigovee_{i}\kappa_i\circ f_i$,
where $f_i=\pproj_i\circ f\colon X\to Y_i$.
Thus, the partial projections $\pproj_i\colon\coprod_{i} Y_i\to Y_i$
are jointly monic.
\myqed
\end{mylemma}

Finally, we give a characterisation of FinPACs.

\begin{mytheorem}[{cf.~\cite[\S5.3]{ArbibM1980}}]
\label{thm:chara-finpac}
A category is a FinPAC if and only if
it has finite coproducts and zero arrows,
and satisfies the following two conditions
for each object $X$:
\par\noindent
\begin{minipage}[t]{.7\textwidth}
\begin{itemize}
\item
the two partial projections
$\pproj_1,\pproj_2\colon X+X\to X$ are jointly monic;
\item
the square on the right is a pullback.
\end{itemize}
\end{minipage}%
\begin{minipage}[t]{.3\textwidth}
\vspace{-4ex}\centering
$
\xymatrix@R-1pc{
(X+X)+X \ar[d]_{\pproj_1}
\ar[r]^-{\nabla+\id} &
X+X \ar[d]^{\pproj_1} \\
X+X\ar[r]^-{\nabla} & X
}
$
\end{minipage}
\end{mytheorem}
\begin{myproof}
For the `if' direction,
we define the partial sum $\ovee$ in the manner of
Proposition~\ref{prop:unique-pcm-enrichment}.
The complete proof is deferred to Appendix~\ref{append:proof-FinPAC}.
\end{myproof}

%=======================================
\section{Partial computation in effectuses and FinPACs with effects}
\label{sec:finpac-with-effects}
%=======================================

Recall from Definition~\ref{def:effectus}
that an effectus $\catB$ has a final object $1$
and finite coproducts $(+,0)$.
Therefore we have the \emph{lift monad} $(-)+1$
on $\catB$.
The unit is the first coprojection $\kappa_1\colon X\to X+1$;
the multiplication is the cotuple
$\cotup{\id_{X+1},\kappa_2}\colon (X+1)+1\to X+1$;
and the Kleisli extension of $f\colon X\to Y+1$
is the cotuple $\cotup{f,\kappa_2}\colon X+1\to Y+1$.
Its Kleisli category, seen as a category for
partial computation, plays an important role in this paper.
Hence we reserve a few notations for it.

We denote by $\KlL{\catB}$
the Kleisli category of the lift monad on $\catB$.
Namely, $\KlL{\catB}$ has the same objects as $\catB$,
and arrows given by $\KlL{\catB}(X,Y)=\catB(X,Y+1)$.
We write $f\colon X\Lto Y$ (`harpoon' arrows) for arrows in $\KlL{\catB}$,
and $g\Lcirc f=\cotup{g,\kappa_2}\circ f$ for the composition in $\KlL{\catB}$.
We denote the canonical functor
$\catB\to\KlL{\catB}$ by $(\Lin{-})$; namely $\Lin{X}=X$ and
$\Lin{f}=\kappa_1\circ f$.
Then $\Lid_X$ denotes the identity $\kappa_1\colon X\Lto X$ in $\KlL{\catB}$.
The Kleisli category $\KlL{\catB}$ has all finite coproducts,
which are inherited from $\catB$
in the way the functor $(\Lin{-})\colon\catB\to\KlL{\catB}$ preserves
the coproducts on the nose.
In other words, a coproduct in $\catBL$ is
a coproduct $\coprod_i X_i$ in $\catB$ with coprojections
$\Lcopr_i\colon X_i\Lto\coprod_i X_i$,
where $\kappa_i$ are coprojections in $\catB$.
For arrows $f,g$ in $\catBL$,
we write $f\Lplus g=[\Lcopr_1\Lcirc f,\Lcopr_2\Lcirc g]$
in order to distinguish it from $f+g$ in $\catB$.
The base category $\catB$ is understood as the `total' part of $\catBL$
via $(\Lin{-})\colon\catB\to\KlL{\catB}$,
see also Proposition~\ref{prop:isom-B-Bot}.

We first collect basic facts on an effectus $\catB$ and
the Kleisli category $\catBL$.

\begin{mylemma}
\label{lem:effectus-basic-facts}
Let $\catB$ be an effectus.
\begin{enumerate}
\item
In $\catB$, coprojections $\kappa_i$ are monic.
Thus, the functor $(\Lin{-})\colon\catB\to\KlL{\catB}$ is faithful.
\item
In $\catB$, squares of the form (K) below
are pullbacks, generalising (K$^=$).
\[
\xymatrix@R-1pc{
A\ar[r]^{g} \ar[d]_{\kappa_1}
\ar@{}[dr]|{\textstyle\text{(K)}} &
B \ar[d]^{\kappa_1} \\
A+X\ar[r]_{g+f} & B+Y
}
\]
\item
$\KlL{\catB}$ has zero arrows
$0_{XY}\colon X\Lto Y$ given by
$X\stackrel{\bang_X}{\longto}1
\stackrel{\kappa_2}{\longto}Y+1$ in $\catB$.
\item
In $\catBL$, the partial projections
$\pproj_1,\pproj_2\colon X+X\Lto X$ are jointly monic.
\item\label{enum:pb-in-catBL}
For a (`total') arrow $f\colon X\to Y$ in $\catB$,
the following square is a pullback in $\catBL$.
\[
\makexypartial
\xymatrix@R-1pc{
X+A \ar[r]^{\Lin{f}\Lplus\Lid}
\ar[d]_{\pproj_1}
& Y+A \ar[d]^{\pproj_1} \\
X \ar[r]^{\Lin{f}} & Y
}
\]
\end{enumerate}
\end{mylemma}
\begin{myproof}
For 1 and 2, see \cite[Lemma~10]{Jacobs2015NewDir}.
\begin{enumerate}\setcounter{enumi}{2}
\item Straightforward.
\auxproof{For $f\colon X\Lto Y$
and $g\colon Z\Lto W$ in $\KlL{\catB}$,
\begin{align*}
(\kappa_2\circ\bang_Y)\Lcirc f
&= \cotup{\kappa_2\circ\bang_Y,\kappa_2}\circ f \\
&= \kappa_2\circ \cotup{\bang_Y,\id_1}\circ f \\
&= \kappa_2\circ \bang_{Y+1}\circ f \\
&= \kappa_2\circ \bang_{X} \\
&= \cotup{g,\kappa_2}\circ \kappa_2\circ \bang_{X} \\
&= g\Lcirc (\kappa_2\circ \bang_{X})
\enspace.
\end{align*}}%
\item
It holds precisely when
$\cotup{\pproj_1,\kappa_2},\cotup{\pproj_2,\kappa_2}
\colon (X+X)+1\to X+1$ are jointly monic in $\catB$,
which is indeed the case; see~\cite[Lemma~11]{Jacobs2015NewDir}.
\item
The square is a pullback in $\catBL$
if and only if the left-hand square below
is pullback in $\catB$.
\[
\xymatrix@C+1pc@R-1pc{
(X+A)+1 \ar[r]^{\cotup{\Lin{f}\Lplus\Lin{\id},\kappa_2}}
\ar[d]_{\cotup{\pproj_1,\kappa_2}}
& (Y+A)+1 \ar[d]^{\cotup{\pproj_1,\kappa_2}} \\
X+1 \ar[r]^{\cotup{\Lin{f},\kappa_2}} & Y+1
}
\qquad
\xymatrix@C+1pc@R-1pc{
X+(A+1) \ar[r]^{f+\id}
\ar[d]_{\id+\bang}
\pb{315}
& Y+(A+1) \ar[d]^{\id+\bang} \\
X+1 \ar[r]^{f+\id} & Y+1
}
\]
Up to isomorphism,
it coincides with the right-hand pullback (E).
\end{enumerate}
\end{myproof}

\begin{mytheorem}
\label{thm:effectus-finpac}
For an effectus $\catB$,
the category $\catBL$ is a FinPAC.
\end{mytheorem}
\begin{myproof}
We use Theorem~\ref{thm:chara-finpac},
a characterisation of a FinPAC.
We have already seen that
$\catBL$ has finite coproducts and zero arrows,
and that the partial projections
$\pproj_1,\pproj_2\colon X+X\Lto X$ are jointly monic.
The required pullback
is an instance of Lemma~\ref{lem:effectus-basic-facts}.\ref{enum:pb-in-catBL}
via $f=\nabla\colon X+X\to X$.
\end{myproof}

Thus, $\catBL$ is $\PCM$-enriched,
and in particular the sets of predicates $\catB(X,1+1)=\catBL(X,1)$
are PCMs. Jacobs showed that predicates have even more structures.

\begin{myproposition}[{\cite[Proposition~13]{Jacobs2015NewDir}}]
\label{prop:effectus-EA}
Let $\catB$ be an effectus.
For each $X\in\catB$,
the homset $\catB(X,1+1)=\KlL{\catB}(X,1)$ is an effect algebra
with the top $1_X\coloneqq\kappa_1\circ\bang_X=\Lin{\bang}_X$.
\myqed
\end{myproposition}

\noindent
Note that the PCM structure on predicates
given by Jacobs~\cite[Definition~12]{Jacobs2015NewDir}
coincides with our partially additive structure
(see Proposition~\ref{prop:unique-pcm-enrichment}).
Crucially, the category $\catBL$ is not only equipped with
both the partially additive and the effect algebra structure,
but satisfies suitable conditions that relate them.
We give a name to such categories, since
they will turn out to characterise $\catBL$.

\begin{mydefinition}
\label{def:finpac-with-effects}
A \emph{FinPAC with effects} is a FinPAC $\catC$
with a special object $I\in\catC$
such that hom-PCMs $\catC(X,I)$ are effect algebras
for all $X\in\catC$,
satisfying the two conditions below.
We write $1_X$ and $0_X$ ($=0_{XI}$) for the top
and the bottom of $\catC(X,I)$.
\begin{enumerate}
\item\label{def-enum:zero-monic}
$1_Y\circ f=0_X$ implies $f=0_{XY}$
for all $f\colon X\to Y$.
\item\label{def-enum:reflect-perp}
$1_Y\circ f\perp 1_Y\circ g$
implies $f\perp g$
for all $f,g\colon X\to Y$.
\end{enumerate}
\end{mydefinition}

\begin{mytheorem}
\label{thm:eff-to-Peff}
Let $\catB$ be an effectus.
Then $(\KlL{\catB},1)$
is a FinPAC with effects.
\end{mytheorem}
\begin{myproof}
We check the two requirement.
\ref{def-enum:zero-monic})
Assume that $1_Y\Lcirc f=0_X$,
i.e.\ $(\bang_Y+\id)\circ f=\kappa_2\circ\bang_X$.
Using a pullback (K) with the symmetry of coproducts,
we obtain $f=\kappa_2\circ\bang_X=0_{XY}$
as in the left diagram below.
\[
\xymatrix@R-1pc{
X \ar@(d,l)[ddr]_{f}
\ar@{-->}[dr]^{\bang}
\ar@(r,u)[drr]^{\bang}
&& \\
&1\ar[d]_{\kappa_2}\ar@{=}[r] \pb{315} &
1\ar[d]^{\kappa_2} \\
&Y+1 \ar[r]^{\bang+\id} & 1+1
}
\qquad
\makexypartial
\xymatrix@R-1pc{
X \ar@(d,l)[ddr]_{f}
\ar@{--^>}[dr]^{c}
\ar@(r,u)[drr]^{b} &\\
&Y+1 \ar[d]_{\pproj_1}
\pb{315}
\ar[r]^-{1\Lplus\Lin{\id}} &
1+1 \ar[d]^{\pproj_1} \\
&Y\ar[r]^-{1} & 1
}\qquad
\xymatrix@R-1pc{
X \ar@(d,l)[ddr]_{g}
\ar@{--^>}[dr]^{d}
\ar@(r,u)[drr]^{c} &\\
&Y+Y \ar[d]_{\pproj_2}
\pb{315}
\ar[r]^-{\Lin{\id}\Lplus 1} &
Y+1 \ar[d]^{\pproj_2} \\
&Y\ar[r]^-{1} & 1
}
\]
\ref{def-enum:reflect-perp}) Let $b\colon X\Lto 1+1$ be a bound for
$1_Y\Lcirc f$ and $1_X\Lcirc g$.
Note that $1_Y=\Lin{\bang}_Y$
is a `total' arrow.
Then we use a pullback of Lemma~\ref{lem:effectus-basic-facts}.\ref{enum:pb-in-catBL}
and obtain a mediating map $c\colon X\Lto Y+1$ as in
the middle diagram above.
Using a similar pullback given by the symmetry of coproducts,
we obtain $d\colon X\Lto Y+Y$ as in the right diagram above.
Then it is straightforward to check $d$ is a bound for $f$ and $g$.
\end{myproof}

In a FinPAC with effects $(\catC,I)$,
we call an arrow $p\colon X\to I$
a \emph{predicate} on $X$,
and write $\Pred(X)=\catC(X,I)$,
which is by definition an effect algebra.
When $\catC=\catBL$,
this definition coincides with predicates in the effectus $\catB$,
since $\catBL(X,1)=\catB(X,1+1)$.
For an arrow $f\colon X\to Y$ in $\catC$,
we call $1_Y\circ f\in\Pred(X)$
the \emph{domain predicate} of $f$
and write $\Dp(f)=1_Y\circ f$.
We then have a PCM-homomorphism
$\Dp\colon\catC(X,Y)\to\Pred(X)$.
We say an arrow $f\colon X\to Y$ in $\catC$
is \emph{total} if $\Dp(f)=1_X$.
It is easy to see that
all objects of $\catC$ with total arrows
form a subcategory of $\catC$, which is denoted by
$\totcat{\catC}$.

\begin{myproposition}
\label{prop:isom-B-Bot}
For an effectus $\catB$,
the functor $(\Lin{-})\colon \catB\to \KlL{\catB}$
restricts to an isomorphism
$\catB\cong\totcat{(\KlL{\catB})}$.
\end{myproposition}
\begin{myproof}
Since $(\Lin{-})$ is faithful,
it suffices to show that
an arrow $f\colon X\Lto Y$ in $\catBL$ is
total if and only if $f=\Lin{g}$ for some
$g\in X\to Y$ in $\catB$.
If $f=\Lin{g}$ then $\Dp(f)=
1_Y\Lcirc f=\Lin{\bang}_Y\Lcirc \Lin{g}
=\Lin{\bang_Y\circ g}=1_X$.
Conversely, assume that $\Dp(f)=1_X$,
i.e.\ $(\bang_Y+\id)\circ f=\kappa_1\circ\bang_X$.
Using a pullback (K),
we obtain $g\colon X\to Y$ with $f=\kappa_1\circ g=\Lin{g}$.
\end{myproof}

We list several basic properties of a FinPAC with effects.

\begin{mylemma}
\label{lem:finpac-w-effect-basics}
In a FinPAC with effects $(\catC,I)$, the following hold.
\begin{enumerate}
\item\label{lem-enum:Dp-zero}
$f=0_{XY}$ if and only if $\Dp(f)=0_X$
for all $f\colon X\to Y$.
\item\label{lem-enum:Dp-perp}
$f_1,\dotsc,f_n$ are orthogonal
if and only if
$\Dp(f_1),\dotsc,\Dp(f_n)$ are orthogonal
for all $f_1,\dotsc,f_n\colon X\to Y$.
In that case, $\Dp(\bigovee_i f_i)=\bigovee_i \Dp(f_i)$.
\item\label{lem-enum:Dp-circ}
$\Dp(g\circ f)=\Dp(g)\circ f\le\Dp(f)$
for all $f\colon X\to Y$ and $g\colon Y\to Z$.
If $g$ is total then $\Dp(g\circ f)=\Dp(f)$.
\item\label{lem-enum:split-mono-total}
Any split mono is total. In particular, any isomorphism is total.
\item\label{lem-enum:copr-total}
Coprojections $\kappa_i$ are split monic and hence total.
\item\label{lem-enum:1-eq-id}
$1_I=\id_I\colon I\to I$.
\end{enumerate}
\end{mylemma}
\begin{myproof}
\begin{enumerate}
\item
The `only if' direction holds because $\Dp$ is homomorphism,
while `if' holds by definition.
\item
The binary case is immediate, like~\ref{lem-enum:Dp-zero}.
Then the $n$-ary case follows by induction.
\item
$\Dp(g\circ f)
=1_Z\circ g\circ f
=\Dp(g)\circ f
\le 1_Y\circ f=\Dp(f)$.
We have equality when $g$ is total.
\item
If $g\circ f=\id$,
then $\Dp(f)\ge\Dp(g\circ f)=\Dp(\id)=1$.
\item
Coprojections are split monic as $\pproj_i\circ\kappa_i=\id$.
\item
Note that $\Dp(\id_I)\perp\Dp(\id_I^\bot)$
and $\Dp(\id_I)=1_I$. Then
$\Dp(\id_I^\bot)=0_I$ and hence $\id_I^\bot=0_{II}=0_I$.
Namely $\id_I=1_I$.
\end{enumerate}
\end{myproof}

In a FinPAC with effects,
we have the decomposition property
(Lemma~\ref{lem:FinPAC-nary-decomp})
as a more explicit bijective correspondence
involving domain predicates.

\begin{mylemma}
\label{lem:P-effectus-bij-corr}
Let $\coprod_{i} Y_i$ be a finite coproduct
in a FinPAC with effects. We have the following bijective correspondence.
\begin{prooftree}
\AxiomC{an arrow $f\colon X\longto \coprod_{i} Y_i$}
\doubleLine
\UnaryInfC{a family $(f_i\colon X\longto Y_i)_{i}$
where $(\Dp(f_i))_{i}$ is orthogonal (so that $\bigovee_{i}\Dp(f_i)\le 1_X$)}
\end{prooftree}
They are related in $f_i=\pproj_i\circ f$
and $f=\bigovee_{i}\kappa_i\circ f_i$.
Moreover one has $\Dp(f)=\bigovee_{i}\Dp(f_i)$.
In particular, $f$ is total if and only if
$\bigovee_{i}\Dp(f_i)=1_X$.
\end{mylemma}
\begin{myproof}
Given $f\colon X\to \coprod_{i}Y_i$,
we have the decomposition
$f=\bigovee_{i}\kappa_i\circ f_i$,
where $f_i=\pproj_i\circ f$.
The family $(\Dp(f_i))_{i}$ is orthogonal
because $\Dp(f_i)=\Dp(\kappa_i\circ f_i)$.
Conversely, if $(\Dp(f_i))_{i}$ is orthogonal
for arrows $f_i\colon X\to Y_i$,
then $(\kappa_i\circ f_i)_i$ is orthogonal.
Hence we have the sum
$f=\bigovee_{i}\kappa_i\circ f_i$.
It is easy to see that the correspondence is bijective.
Finally,
$\Dp(f)
=\bigovee_{i}\Dp(\kappa_i\circ f_i)
=\bigovee_{i}\Dp(f_i)$.
\end{myproof}

\begin{mylemma}
\label{lem:P-effectus-coproduct}
Let $(\catC,I)$ be a FinPAC with effects.
Coproducts $\coprod_i X_i$
in $\catC$ restrict to $\totcat{\catC}$,
so that $\totcat{\catC}$ has all finite coproducts.
Moreover, $I$ is final in $\totcat{\catC}$,
and we have an isomorphism $\KlL{(\totcat{\catC})}\cong\catC$,
which is identity on objects, and
sends an arrow $f\colon X\to Y+I$ to $\pproj_1\circ f\colon X\to Y$.
\end{mylemma}
\begin{myproof}
Since coprojections are total,
the coproduct diagram is in $\totcat{\catC}$.
Let $f_i\colon X_i\to Y$ be total arrows.
Then the cotuple $\cotup{f_i}_i\colon \coprod_i X_i\to Y$
is total,
i.e.\ $\Dp(\cotup{f_i}_i)=1_{\coprod_i X_i}$ since
$
\Dp(\cotup{f_i}_i)\circ\kappa_i
=\Dp(\cotup{f_i}_i\circ\kappa_i)
=\Dp(f_i)
=1_{X_i}
=\Dp(\kappa_i)
=1_{\coprod_i X_i}\circ\kappa_i
$
for all $i$.
The object $I$ is final in $\totcat{\catC}$,
because $\totcat{\catC}(X,I)=\{1_X\}$.
%\newpage%adjust

It is easy to see the mapping $f\mapsto \pproj_1\circ f$
is functorial.
To prove $\KlL{(\totcat{\catC})}\cong\catC$,
it suffices to show the functor is full and faithful.
Let $f\in\catC(X,Y)$.
Using Lemma~\ref{lem:P-effectus-bij-corr},
we obtain $g\colon X\to Y+I$
by $g=\kappa_1\circ f\ovee \kappa_2\circ \Dp(f)^\bot$.
Then $g$ is total and $\pproj_1\circ g=f$.
Suppose that $h\colon X\to Y+I$ is a total arrow with
$\pproj_1\circ h=f$.
Consider the decomposition $h=\kappa_1\circ h_1\ovee \kappa_2\circ h_2$,
with $h_1=\pproj_1\circ h$ and $h_2=\pproj_2\circ h$.
We have $h_1=f$ and
$1=\Dp(h_1)\ovee\Dp(h_2)=\Dp(f)\ovee h_2$,
so that $h_2=\Dp(f)^\bot$. Hence $h=g$.
\end{myproof}

\begin{mytheorem}
\label{thm:Peff-to-eff}
Let $(\catC,I)$ be a FinPAC with effects.
The subcategory $\totcat{\catC}$ is an effectus.
\end{mytheorem}
\begin{myproof}
We have already seen that $\totcat{\catC}$
has finite coproducts $(+,0)$ and
a final object $I$.
The joint monicity requirement
is equivalent to say that $\pproj_1,\pproj_2\colon I+I\Lto I$
are jointly monic in $\KlL{(\totcat{\catC})}$,
which is true since $\KlL{(\totcat{\catC})}\cong\catC$.
\auxproof{We show the pullback requirement:
for total arrows $f\colon X\to Y$
and $g\colon A\to B$,
the following diagrams are pullbacks in $\totcat{\catC}$.
\[
\xymatrix@R-1pc{
A+X\ar[r]^{\id+f} \ar[d]_{g+\id}
& A+Y \ar[d]^{g+\id} \\
B+X\ar[r]^{\id+f} & B+Y
}
\qquad
\xymatrix@R-1pc{
A\ar@{=}[r] \ar[d]_{\kappa_1}
& A \ar[d]^{\kappa_1} \\
A+X\ar[r]^{\id+f} & A+Y
}
\]}%
We prove that the squares (E) and (K$^=$) are pullbacks
in $\totcat{\catC}$.

\textbf{(E)}
Let $\alpha\colon Z\to A+Y$ and $\beta\colon Z\to B+X$ be total arrows
with $(g+\id)\circ \alpha=(\id+f)\circ \beta$.
By postcomposing partial projections
$\pproj_i$ to $h\coloneqq(g+\id)\circ \alpha=(\id+f)\circ \beta$,
we obtain
$h_1=g\circ \alpha_1=\beta_1$ and
$h_2=\alpha_2=f\circ \beta_2$,
where $\alpha_i=\pproj_i\circ\alpha$,
$\beta_i=\pproj_i\circ\beta$ and
$h_i=\pproj_i\circ h$.
Note that
$\Dp(\alpha_1)=\Dp(g\circ \alpha_1)=\Dp(h_1)$ and
$\Dp(\beta_2)=\Dp(f\circ\beta_2)=\Dp(h_2)$.
By Lemma~\ref{lem:P-effectus-bij-corr},
we can define a total arrow $\gamma\colon Z\to A+X$ by
$\gamma=\kappa_1\circ\alpha_1\ovee
\kappa_2\circ\beta_2$.
As desired,
$(\id+f)\circ\gamma
= \kappa_1\circ\alpha_1\ovee
\kappa_2\circ f\circ\beta_2
= \kappa_1\circ\alpha_1\ovee
\kappa_2\circ\alpha_2
= \alpha$,
and $(g+\id)\circ\gamma=\beta$ similarly.
To see the uniqueness,
assume that $\gamma\colon Z\to A+X$ satisfies
$(\id+f)\circ\gamma=\alpha$ and $(g+\id)\circ\gamma=\beta$.
Then one has $\pproj_1\circ\gamma=\alpha_1$
and $\pproj_2\circ\gamma=\alpha_2$.
Hence the joint monicity of partial projections implies the uniqueness.

\textbf{(K$^=$)}
Let $\alpha\colon Z\to A$ and $\beta\colon Z\to A+X$
be total arrows with $\kappa_1\circ \alpha=(\id+f)\circ \beta$.
By postcomposing partial projections $\pproj_i$ to
$\kappa_1\circ\alpha=(\id+f)\circ\beta$,
we obtain $\alpha=\beta_1$ and $0_{}=f\circ \beta_2$,
where $\beta_i=\pproj_i\circ\beta$.
Then $\beta_2=0$ since $\Dp(\beta_2)=\Dp(f\circ\beta_2)=0$.
Now we have
$
\beta=\kappa_1\circ\beta_1\ovee\kappa_2\circ\beta_2
=\kappa_1\circ\alpha
$
as desired.
\end{myproof}

\begin{myexample}
Recall from \S{}\ref{subsec:effectus} three examples of effectuses
$(\Cstar_\PU)^\op$, $\Set$ and $\Kl(\Dst)$.
By Theorem~\ref{thm:eff-to-Peff},
the Kleisli categories of
the lift monads on these effectuses are FinPACs with effects
(see also Table~\ref{table:eff-P-eff} in \S{}\ref{sec:intro}).

\vspace{-1ex}
\parpic[r]{%
\begin{minipage}{.3\linewidth}
\begin{prooftree}
\AxiomC{$f\colon A\times\C\longto B$\quad PU-map}
\doubleLine
\UnaryInfC{$g\colon A\longto B$\quad PSU-map}
\end{prooftree}
\end{minipage}}
The Kleisli category $\KlL{((\Cstar_\PU)^\op)}$
is isomorphic to the opposite $(\Cstar_\PSU)^\op$
of the category of $C^*$-algebras and positive
subunital\footnote{%
A positive map $g\colon A\to B$ is said to be \emph{subunital}
if $g(1)\le 1$.} (PSU) maps.
Indeed, we have the bijective correspondence
shown on the right, via
$g(x)=f(x,0)$ and $f(x,\lambda)=g(x)+\lambda(1-g(1))$.
Predicates are PSU-maps $\C\to A$,
which are easily identified with effects
$p\in[0,1]_A\coloneqq\{p\in A\mid 0\le p\le 1\}$.
Then the domain predicate of a PSU-map $f\colon A\to B$
is identified with $f(1)\in[0,1]_B$.
By Lemma~\ref{lem:finpac-w-effect-basics}.\ref{lem-enum:Dp-perp},
the sum $f\ovee g$ of PSU-maps $f,g\colon A\to B$
is defined precisely when $f(1)\perp g(1)$ in $[0,1]_B$,
namely $f(1)+g(1)\le 1$. In that case the sum is defined pointwise:
$(f\ovee g)(x)=f(x)+g(x)$.
Note that $C^*$-algebras with completely positive subunital maps
and $W^*$-algebras with normal (completely) positive subunital maps
work in exactly the same way. The latter is especially important
for semantics of quantum programming languages~\cite{Rennela2014,Cho2014QPL}.

For the classical example,
it is well-known that $\KlL{\Set}\cong\Pfn$, where $\Pfn$
is the category of sets and partial functions.
The domain predicate $\Dp(f)$ of a partial function $f\colon X\rightharpoonup Y$
is identified with its domain of definition $\dom(f)\subseteq X$.
The sum of $f,g\colon X\rightharpoonup Y$
is defined precisely if $\dom(f)\perp\dom(g)$ in $\Pow(X)$,
i.e.\ $\dom(f)\cap\dom(g)=\emptyset$.
In that case $f\ovee g$ is defined on $\dom(f)\cup\dom(g)$
in an obvious way.

For the probabilistic example, we have
$\KlL{\Kl(\Dst)}\cong\Kl(\Dsub)$, where $\Dsub=\Dsub_{[0,1]}$
is the subdistribution monad over $[0,1]$.
This is due to the natural bijections $\Dsub(X)\cong\Dst(X+1)$.
The domain predicate $\Dp(f)$ of a map $f\colon X\to Y$ in $\Kl(\Dsub)$,
i.e.\ a function $f\colon X\to \Dsub(Y)$ is identified with
the `fuzzy' predicate $p\in [0,1]^X$ given by
$p(x)=\sum_y f(x)(y)$.
The sum of functions $f,g\colon X\to \Dsub Y$ is defined
if and only if $\sum_y f(-)(y)\perp \sum_y g(-)(y)$
in $[0,1]^X$, that is,
$\sum_y f(x)(y)+\sum_yg(x)(y)\le 1$ for all $x\in X$.
In that case, the sum is defined by $(f\ovee g)(x)(y)=f(x)(y)+g(x)(y)$.
\end{myexample}

%-----------------------------------------------------------
\section{Categorical equivalence of effectuses and FinPACs with effects}
\label{sec:equivalence-eff-FPE}
%-----------------------------------------------------------

The results in the previous section
are summarised as follows.
For an effectus $\catB$,
the category $\KlL{\catB}$
with $1\in\KlL{\catB}$ is a FinPAC with effects;
and for a FinPAC with effects $(\catC,I)$,
the subcategory $\totcat{\catC}$ is an effectus.
Moreover we have isomorphisms
$\catB\cong\totcat{(\KlL{\catB})}$
and $\catC\cong\KlL{(\totcat{\catC})}$.
We can immediately obtain a characterisation of effectuses.
\begin{mycorollary}
A category $\catB$ is an effectus
if and only if
there is a FinPAC with effects $(\catC,I)$
such that $\catB\cong\totcat{\catC}$.
\myqed
\end{mycorollary}

The results are most naturally presented in terms of
(2-)categorical equivalence.

\begin{mydefinition}
We define a (strict) 2-category $\Eff$ of effectuses as follows.
An object is an effectus $\catB$.
An arrow $F\colon \catA\to \catB$ is a functor that preserves the final object and
finite coproducts.
A 2-cell $\alpha\colon F\To G$ is a natural transformation
that is monoidal w.r.t.\ $(+,0)$.
We also define a 2-category $\FPE$ of FinPACs with effects
as follows.
An object is a FinPAC with effects $(\catC,I)$.
An arrow $F\colon (\catC,I_\catC)\to(\catD,I_\catD)$
is a functor $F\colon\catC\to\catD$
that preserves finite coproducts
and ``preserves the truth'' in the sense that
$1_{FI_\catC}\colon FI_\catC\to I_\catD$ is an isomorphism,
and $1_{FI}\circ F1_X=1_{FX}$ for all $X\in\catC$.
A 2-cell $\alpha\colon F\To G$ is a natural transformation
that is monoidal w.r.t.\ $(+,0)$, and satisfies
$1_{GI}\circ\alpha_I=1_{FI}$.
\end{mydefinition}

\begin{mytheorem}
The assignments $\catB\mapsto\KlL{\catB}$
and $\catC\mapsto\totcat{\catC}$ extend to
2-functors $\KlL{(-)}\colon\Eff\to\FPE$
and $\totcat{(-)}\colon\FPE\to\Eff$ respectively.
Moreover, they form a 2-equivalence of 2-categories $\Eff\simeq\FPE$.
\end{mytheorem}
\begin{myproof}
The essential part is already done.
The rest, checking functoriality and naturality,
is mostly routine.
We defer the details to
Appendix~\ref{append:equiv-of-effectus-P-effectus}.
\end{myproof}

%===================================================
\section{State-and-effect triangles over FinPACs with effects}
\label{sec:triangle-P-effectus}
%===================================================

Let $(\catC,I)$ be a FinPAC with effects.
Recall that $\Pred(X)=\catC(X,I)$
is the set of predicates on $X$.
We call an arrow $\omega\colon I\to X$
a \emph{substate} on $X$,
an arrow $r\colon I\to I$
a \emph{scalar}.
We write $\SStat(X)=\catC(I,X)$
for the set of substates on $X$,
and let $M=\catC(I,I)$ be the set of scalars.

\begin{myproposition}
Let $(\catC,I)$ be a FinPAC with effects.
\begin{enumerate}
\item
The effect algebra $M=\catC(I,I)$ is
an effect monoid with the composition $\circ$
as a multiplication.
\item
For each $X\in\catC$,
the effect algebra $\Pred(X)=\catC(X,I)$ is
an effect module over $M$,
with the composition $\circ$ as a scalar multiplication.
\item
For each $X\in\catC$,
the PCM $\SStat(X)=\catC(I,X)$ is
a PCMod over $M$
with the composition $\circ$ as a (right) scalar multiplication.
Moreover it is subconvex.
\end{enumerate}
\end{myproposition}
\begin{myproof}
Straightforward, but note that $1_I=\id_I\in M$. To see
$\SStat(X)$ is subconvex, use $\Dp(\omega\circ r)\le r$.
\end{myproof}

Combining the dual adjunction
$(\GEMod_M)^\op\rightleftarrows\SConv_M$
from Proposition~\ref{prop:emod-conv-duality},
we obtain a state-and-effect triangle.
We use the category $\GEMod_M$ of GEMod's
because induced predicate transformers
do not necessarily preserve the truth predicates.

\noindent
\begin{minipage}[t]{.65\textwidth}
\begin{mytheorem}
For a FinPAC with effects $(\catC,I)$,
the hom-functors $\catC(-,I)$
and $\catC(I,-)$ give rise to the functors in
the diagram on the right,
constituting a state-and-effect triangle.
\end{mytheorem}
\end{minipage}
\begin{minipage}[t]{.35\textwidth}
\vspace{.5ex}\centering
$
\xymatrix@C=1pc@R=1.5pc{
(\GEMod_M)^\op \ar@/^1.5ex/[rr]
\ar@{}[rr]|-{\top}&
&
\SConv_M \ar@/^1.5ex/[ll]
\\
&
\catC\ar[ul]^{\catC(-,I)\,=\,\Pred\;\;}
\ar[ur]_{\SStat\,=\,\catC(I,-)} &
}
$
\vspace{-1ex}
\end{minipage}
\begin{myproof}
It is easy to check that the precomposition $\catC(f,I)$
and the postcomposition $\catC(I,f)$ are desired homomorphisms.
\end{myproof}

Examples of this type of state-and-effect triangles
have already appeared in~\cite{JacobsCMCS2014,Rennela2013Master},
but the general construction is new.
Substates in the quantum example $(\Cstar_\PSU)^\op$
are PSU-maps $\omega\colon A\to\C$.
In the classical example $\Pfn$,
substates on $X$ are either elements $x\in X$ or the `bottom'.
In the probabilistic example $\Kl(\Dsub)$,
substates are subdistributions $\omega\in\Dsub(X)$.

As is the case for effectuses (\S\ref{subsec:effectus}),
there is an \emph{abstract Born rule} given by
$(\omega\vDash p)\coloneqq p\circ \omega\in M$ for $\omega\colon I\to X$
and $p\colon X\to I$. The map $\vDash\colon\SStat(X)\times\Pred(X)\to M$
is an appropriate bihomomorphism, so that by ``currying'', we obtain the following maps
$\alpha_X$ and $\beta_X$ in the bijective correspondence of the dual adjunction.

\begin{prooftree}
\AxiomC{$\alpha_X\colon\Pred(X)\longto\SConv_M(\SStat(X),M)$ in $\GEMod_M$}
\doubleLine
\UnaryInfC{$\beta_X\colon\SStat(X)\longto\GEMod_M(\Pred(X),M)$ in $\SConv_M$}
\end{prooftree}

\noindent These maps $\alpha$ and $\beta$ give
natural transformations which fill the state-and-effect triangle

In a FinPAC with effects,
a \emph{state} on $X$ is
a substate $\omega\colon I\to X$ with $\Dp(\omega)=1$
(i.e.\ a total substate),
and the set of states is denoted by
$\Stat(X)=\totcat{\catC}(I,X)$.
This definition accords with states in an effectus,
since $\totcat{(\KlL{\catB})}(1,X)\cong\catB(1,X)$.
The set $\Stat(X)$ is a subset of $\SStat(X)$ that is closed under
convex sum, hence $\Stat(X)$ is a convex set, giving a functor
$\Stat\colon\totcat{\catC}\to\Conv_M$.
On the other hand, we obtain a functor
$\Pred\colon\totcat{\catC}\to(\EMod_M)^\op$
as a restriction of $\Pred\colon\catC\to(\GEMod_M)^\op$,
since predicate transformers induced by total arrows
preserve the truth predicates.
This is an alternative way to obtain a state-and-effect triangle
over an effectus shown in Figure~\ref{fig:state-and-effect-triangle}
(cf.\ \cite{Jacobs2015NewDir}).

In what follows,
we will focus on a FinPAC with effects satisfying
`normalisation' (of states).
A FinPAC with effects $(\catC,I)$
satisfies \emph{normalisation} if
for each object $X$ and
for each substate $\omega\in\SStat(X)$
that is nonzero ($\omega\ne 0_{IX}$),
there exists a unique state $\tilde{\omega}\in\Stat(X)$
such that $\omega=\tilde{\omega}\circ\Dp(\omega)$.
An effectus $\catB$ satisfies \emph{normalisation}
if the corresponding FinPAC with effects
$(\KlL{\catB},1)$ satisfies normalisation.
An effectus with normalisation was
introduced and studied in~\cite{JacobsWW2015}, where
most results are restricted
to the case when the set of scalars $M$ is the unit interval $[0,1]$.
In fact, if an effectus or FinPAC with effects
satisfies normalisation,
then the scalars are already `good' enough
to take away the restriction $M=[0,1]$.

\begin{mydefinition}
\label{def:division-emon}
An effect monoid $M$ has \emph{division} if
for all $s,t\in M$ with
$s\le t$ and $t\ne 0$,
there exists unique `quotient' $q\in M$ such that
$q\cdot t=s$.
The quotient $q$ is denoted by $s/t$.
We call such an effect monoid
a \emph{division effect monoid}.
\end{mydefinition}

\begin{myproposition}
If a FinPAC with effects $(\catC,I)$ satisfies normalisation,
then the effect monoid of scalars $M=\catC(I,I)$
has division.
\end{myproposition}
\begin{myproof}
Let $s,t\in M$ be scalars with $s\le t$ and $t\ne 0$.
Let $s'=t\ominus s$, so that $s\ovee s'=t$.
Let $\omega=\kappa_1\circ s\ovee \kappa_2\circ s'\colon I\to I+I$,
which is nonzero because $\Dp(\omega)=s\ovee s'=t\ne 0$.
By normalisation there is a state $\tilde{\omega}\colon I\to I+I$
with $\omega=\tilde{\omega}\circ\Dp(\omega)=\tilde{\omega}\circ t$.
Then $s=\pproj_1\circ\omega=\pproj_1\circ\tilde{\omega}\circ t$.
Therefore $\pproj_1\circ\tilde{\omega}$ is a desired quotient.
To see the uniqueness of the quotient, assume
that $q\in M$ satisfies $s=q\circ t$.
Then $s'=t\ominus s=t\ominus(q\circ t)=q^\bot\circ t$.
Let $\omega_q=\kappa_1\circ q\ovee\kappa_2\circ q^\bot\colon I\to I+I$,
which is a state and
$
\omega_q\circ t=
\kappa_1\circ q\circ t\ovee
\kappa_2\circ q^\bot\circ t
=\kappa_1\circ s\ovee
\kappa_2\circ s'=\omega
$.
By the uniqueness of normalisation,
we obtain $\omega_q=\tilde{\omega}$.
Therefore $\pproj_1\circ\tilde{\omega}=\pproj_1\circ\omega_q=q$.
\end{myproof}

The division indeed satisfies desired properties,
see Lemmas~\ref{lem:div-emon-cancel} and~\ref{lem:div-emon-hom}.
It allows us to obtain the following result, by
generalising $M=[0,1]$ to any division effect monoid.

\begin{mytheorem}[{\cite[Corollary~19]{JacobsWW2015}}]
\label{thm:eff-normalis-state-and-effect}
Let $\catB$ be an effectus satisfying normalisation.
Then, all the categories and the functors
in the state-and-effect triangle over $\catB$
(Figure~\ref{fig:state-and-effect-triangle})
are objects and arrows in $\Eff$.
\qed
\end{mytheorem}

\noindent
Note that, unlike \cite{JacobsWW2015},
we simply use $\Conv_M$ rather than
the category of \emph{cancellative} convex sets.
This is because we use a weaker variant of the joint monicity requirement
in Definition~\ref{def:effectus},
and $\Conv_M$ is indeed an effectus in our sense;
see Proposition~\ref{prop:conv-effectus}.
Furthermore,
it is straightforward to check the following.

\begin{mylemma}
Let $M$ be a division effect monoid.
The unit and the counit of the adjunction
$(\EMod_M)^\op\rightleftarrows\Conv_M$
are 2-cells in $\Eff$.
Namely, $(\EMod_M)^\op\rightleftarrows\Conv_M$
is an adjunction in the 2-category $\Eff$.
\myqed
\end{mylemma}

\noindent
In the light of the 2-equivalence $\Eff\simeq\FPE$,
we obtain a corresponding state-and-effect triangle
over a FinPAC with effects.

\noindent
\begin{minipage}[t]{.62\textwidth}
\begin{mycorollary}
Let $(\catC,I)$ be a FinPAC with effects satisfying
normalisation.
We have a state-and-effect triangle on the right,
where the categories, the functors
and the adjunction are in $\FPE$.
\end{mycorollary}
\end{minipage}
\begin{minipage}[t]{.38\textwidth}
\vspace{0ex}\centering
$
\xymatrix@C=0pc@R=1.5pc{
\KlL{((\EMod_M)^\op)} \ar@/^1.5ex/[rr]
\ar@{}[rr]|-{\top}&
& \KlL{(\Conv_M)} \ar@/^1.5ex/[ll]
\\
&
\KlL{(\totcat{\catC})}
\mathrlap{{}\cong\catC}
\ar[ul]^{\KlL{\Pred}}
\ar[ur]_{\KlL{\Stat}} &
}
$
\end{minipage}
\par\vspace{-2.5ex}
\hfill\ensuremath{\blacksquare}

%====================
\section{Conclusions}
%====================

We studied partial computation in effectuses,
giving a fundamental equivalence of effectuses and FinPACs with effects.
Despite the equivalence,
FinPACs with effects sometimes have an advantage over effectuses,
because they have richer structures such as the finitely partially additive structure.
For instance, an \emph{instrument} map $\instr_p\colon X\to X+\dotsb+X$
for an `$n$-test' $p\colon X\to 1+\dotsb+1$ in an effectus
allow us to perform a (quantum) measurement, with $n$
outcomes~\cite[Assumption~2]{Jacobs2015NewDir}.
Switching to a FinPAC with effects, we can decompose such an instrument map
to $n$ `partial' endomaps $X\to X$, which give a simpler formulation.
The details will be elaborated in a subsequent paper.

Recently the author and his colleagues studied
\emph{quotient--comprehension chains}~\cite{ChoJWW2015QC}
which are related to such instrument maps and measurement. It
is worth noting that many examples of quotient--comprehension
chains are given by FinPACs with effects, including a quantum setting
via $W^*$-algebras.
An important future work
is thus to give a categorical axiomatisation of such a quotient--comprehension chain
in the effectus / `FinPAC with effects' framework.

\section*{Acknowledgements}

I would like to thank
Bart Jacobs for suggesting the problem to characterise the Kleisli
category of the lift monad on an effectus;
Bas Westerbaan for fruitful discussions to solve it;
and Robin Adams for helpful comments.

%\bibliographystyle{eptcs}
%\bibliography{jabref}
%\bibliography{local}

\begin{thebibliography}{10}
\providecommand{\bibitemdeclare}[2]{}
\providecommand{\surnamestart}{}
\providecommand{\surnameend}{}
\providecommand{\urlprefix}{Available at }
\providecommand{\url}[1]{\texttt{#1}}
\providecommand{\href}[2]{\texttt{#2}}
\providecommand{\urlalt}[2]{\href{#1}{#2}}
\providecommand{\doi}[1]{doi:\urlalt{http://dx.doi.org/#1}{#1}}
\providecommand{\bibinfo}[2]{#2}

\bibitemdeclare{article}{ArbibM1980}
\bibitem{ArbibM1980}
\bibinfo{author}{Michael~A. \surnamestart Arbib\surnameend} \&
  \bibinfo{author}{Ernest~G. \surnamestart Manes\surnameend}
  (\bibinfo{year}{1980}): \emph{\bibinfo{title}{Partially additive categories
  and flow-diagram semantics}}.
\newblock {\sl \bibinfo{journal}{Journal of Algebra}}
  \bibinfo{volume}{62}(\bibinfo{number}{1}), pp. \bibinfo{pages}{203--227},
  \doi{10.1016/0021-8693(80)90212-4}.

\bibitemdeclare{inproceedings}{Cho2014QPL}
\bibitem{Cho2014QPL}
\bibinfo{author}{Kenta \surnamestart Cho\surnameend} (\bibinfo{year}{2014}):
  \emph{\bibinfo{title}{Semantics for a Quantum Programming Language by
  Operator Algebras}}.
\newblock In: {\sl \bibinfo{booktitle}{QPL 2014}}, {\sl
  \bibinfo{series}{EPTCS}} \bibinfo{volume}{172}, pp.
  \bibinfo{pages}{165--190}, \doi{10.4204/EPTCS.172.12}.

\bibitemdeclare{inproceedings}{ChoJWW2015QC}
\bibitem{ChoJWW2015QC}
\bibinfo{author}{Kenta \surnamestart Cho\surnameend}, \bibinfo{author}{Bart
  \surnamestart Jacobs\surnameend}, \bibinfo{author}{Bas \surnamestart
  Westerbaan\surnameend} \& \bibinfo{author}{Bram \surnamestart
  Westerbaan\surnameend} (\bibinfo{year}{2015}):
  \emph{\bibinfo{title}{Quotient--Comprehension Chains}}.
\newblock In: {\sl \bibinfo{booktitle}{{QPL} 2015}}.
\newblock \bibinfo{note}{To appear}.

\bibitemdeclare{inproceedings}{JacobsCMCS2014}
\bibitem{JacobsCMCS2014}
\bibinfo{author}{Bart \surnamestart Jacobs\surnameend} (\bibinfo{year}{2014}):
  \emph{\bibinfo{title}{Dijkstra Monads in Monadic Computation}}.
\newblock In: {\sl \bibinfo{booktitle}{CMCS 2014}}, {\sl
  \bibinfo{series}{LNCS}} \bibinfo{volume}{8446},
  \bibinfo{publisher}{Springer}, pp. \bibinfo{pages}{135--150},
  \doi{10.1007/978-3-662-44124-4_8}.

\bibitemdeclare{article}{Jacobs2015NewDir}
\bibitem{Jacobs2015NewDir}
\bibinfo{author}{Bart \surnamestart Jacobs\surnameend} (\bibinfo{year}{2015}):
  \emph{\bibinfo{title}{New Directions in Categorical Logic, for Classical,
  Probabilistic and Quantum Logic}}.
\newblock {\sl \bibinfo{journal}{{LMCS}}}.
\newblock \bibinfo{note}{To appear. \newblock {arXiv}:1205.3940v4 [math.LO]}.

\bibitemdeclare{article}{JacobsM2012Coref}
\bibitem{JacobsM2012Coref}
\bibinfo{author}{Bart \surnamestart Jacobs\surnameend} \&
  \bibinfo{author}{Jorik \surnamestart Mandemaker\surnameend}
  (\bibinfo{year}{2012}): \emph{\bibinfo{title}{Coreflections in Algebraic
  Quantum Logic}}.
\newblock {\sl \bibinfo{journal}{Foundations of Physics}}
  \bibinfo{volume}{42}(\bibinfo{number}{7}), pp. \bibinfo{pages}{932--958},
  \doi{10.1007/s10701-012-9654-8}.

\bibitemdeclare{inproceedings}{JacobsWW2015}
\bibitem{JacobsWW2015}
\bibinfo{author}{Bart \surnamestart Jacobs\surnameend}, \bibinfo{author}{Bas
  \surnamestart Westerbaan\surnameend} \& \bibinfo{author}{Bram \surnamestart
  Westerbaan\surnameend} (\bibinfo{year}{2015}): \emph{\bibinfo{title}{States
  of Convex Sets}}.
\newblock In: {\sl \bibinfo{booktitle}{FoSSaCS 2015}}, {\sl
  \bibinfo{series}{LNCS}} \bibinfo{volume}{9034},
  \bibinfo{publisher}{Springer}, pp. \bibinfo{pages}{87--101},
  \doi{10.1007/978-3-662-46678-0_6}.

\bibitemdeclare{book}{ManesA1986}
\bibitem{ManesA1986}
\bibinfo{author}{Ernest~G. \surnamestart Manes\surnameend} \&
  \bibinfo{author}{Michael~A. \surnamestart Arbib\surnameend}
  (\bibinfo{year}{1986}): \emph{\bibinfo{title}{Algebraic Approaches to Program
  Semantics}}.
\newblock \bibinfo{series}{Monographs in Computer Science},
  \bibinfo{publisher}{Springer}, \doi{10.1007/978-1-4612-4962-7}.

\bibitemdeclare{article}{Moggi1991}
\bibitem{Moggi1991}
\bibinfo{author}{Eugenio \surnamestart Moggi\surnameend}
  (\bibinfo{year}{1991}): \emph{\bibinfo{title}{Notions of computation and
  monads}}.
\newblock {\sl \bibinfo{journal}{Information and Computation}}
  \bibinfo{volume}{93}(\bibinfo{number}{1}), pp. \bibinfo{pages}{55--92},
  \doi{10.1016/0890-5401(91)90052-4}.

\bibitemdeclare{mastersthesis}{Rennela2013Master}
\bibitem{Rennela2013Master}
\bibinfo{author}{Mathys \surnamestart Rennela\surnameend}
  (\bibinfo{year}{2013}): \emph{\bibinfo{title}{On operator algebras in quantum
  computation}}.
\newblock Master's thesis, \bibinfo{school}{Universit{\'e} Paris 7 Denis
  Diderot}.

\bibitemdeclare{inproceedings}{Rennela2014}
\bibitem{Rennela2014}
\bibinfo{author}{Mathys \surnamestart Rennela\surnameend}
  (\bibinfo{year}{2014}): \emph{\bibinfo{title}{Towards a Quantum Domain
  Theory: Order-enrichment and Fixpoints in {W}*-algebras}}.
\newblock In: {\sl \bibinfo{booktitle}{{MFPS} {XXX}}}, {\sl
  \bibinfo{series}{ENTCS}} \bibinfo{volume}{308},
  \bibinfo{publisher}{Elsevier}, pp. \bibinfo{pages}{289--307},
  \doi{10.1016/j.entcs.2014.10.016}.

\end{thebibliography}

%////////////////////////////
\appendix
%////////////////////////////

%==================================
\section{Omitted proofs in Section~\ref{sec:FinPAC}}
\label{append:proof-FinPAC}
%==================================

We write $\nset{n}=\{1,\dotsc,n\}$
for the $n$ element set, and
$n\cdot X=\coprod_{i\in\nset{n}} X$
for an $n$-fold coproduct.

\begin{mylemma}
\label{lem:n-ary-finpac-axioms}
In a FinPAC the following hold.
\begin{enumerate}
\item\label{prop-enum:fin-compat}
A family $(f_i\colon X\to Y)_{i\in\nset{n}}$
is orthogonal whenever $(f_i)_{i\in\nset{n}}$ is compatible
in the sense that there exists a `bound'
$b\colon X\to n\cdot Y$ such that $f_i=\pproj_i\circ b$.
\item\label{prop-enum:fin-untying}
If a family $(f_i\colon X\to Y)_{i\in\nset{n}}$ is orthogonal,
then a family $(\kappa_i\circ f_i\colon X\to n\cdot Y)_{i\in\nset{n}}$
is orthogonal too.
\end{enumerate}
\end{mylemma}

\begin{myproof}
\textbf{\ref{prop-enum:fin-compat}.}
We prove the following stronger statement by induction on $n$.
\begin{itemize}
\item\itshape
If a family $(f_i\colon X\to Y)_{i\in\nset{n}}$
is compatible via a bound $b\colon X\to n\cdot Y$,
then it is orthogonal and $\bigovee_{i\in\nset{n}}f_i=\nabla\circ b$.
\end{itemize}
The base case ($n=0$) is trivial.
To show the induction step,
let $(f_i\colon X\to Y)_{i\in\nset{n+1}}$ be a compatible family
via a bound $b\colon X\to (n+1)\cdot Y$.
Let $\alpha\colon (n+1)\cdot Y\to n\cdot Y+Y$ be the canonical
associativity isomorphism.
Then it is easy to see that
$(f_i)_{i\in\nset{n}}$ is compatible via
$\pproj_1\circ\alpha\circ b\colon X\to n\cdot Y$.
By the induction hypothesis $(f_i)_{i\in\nset{n}}$
is orthogonal and $\bigovee_{i\in\nset{n}} f_i=\nabla\circ\pproj_1\circ\alpha\circ b$.
Note that
\begin{align*}
\pproj_1\circ(\nabla+\id)\circ\alpha\circ b
&= \nabla\circ\pproj_1\circ\alpha\circ b
= \bigovee_{i\in\nset{n}} f_i \\
\pproj_2\circ(\nabla+\id)\circ\alpha\circ b
&= \pproj_2\circ\alpha\circ b
= \pproj_{n+1}\circ b
= f_{n+1}
\enspace.
\end{align*}
Therefore $\bigovee_{i\in\nset{n}} f_i
\perp f_{n+1}$ via
$(\nabla+\id)\circ\alpha\circ b$,
so that $(f_i)_{i\in\nset{n+1}}$ is orthogonal.
Moreover we have
\[
\bigovee_{i\in\nset{n+1}} f_i
=\Bigl(\bigovee_{i\in\nset{n}} f_i\Bigr)\ovee f_{n+1}
=\nabla\circ (\nabla+\id)\circ\alpha\circ b
=\nabla\circ b
\enspace.
\]

\textbf{\ref{prop-enum:fin-untying}.}
We prove it by induction on $n$.
The base case $n=0$ is trivial.
Let $(f_i\colon X\to Y)_{i\in\nset{n+1}}$ be an orthogonal family.
Then $n$ arrows $f_1\ovee f_{n+1},f_2,\dotsc,f_n$ are orthogonal.
By the induction hypothesis,
\[
\kappa_1\circ(f_1\ovee f_{n+1})=\kappa_1\circ f_1\ovee \kappa_1\circ f_{n+1},
\kappa_2\circ f_2,\dotsc,\kappa_n\circ f_n
\colon X\to n\cdot Y
\]
are orthogonal.
This implies that $\bigovee_{i\in\nset{n}}\kappa_i\circ f_i$
and $\kappa_1\circ f_{n+1}$ are orthogonal.
By the untying axiom,
$\kappa_1\circ \bigovee_{i\in\nset{n}}\kappa_i\circ f_i=
\bigovee_{i\in\nset{n}}\kappa_1\circ\kappa_i\circ f_i$
and $\kappa_2\circ \kappa_1\circ f_{n+1}$
are orthogonal.
It follows that
\[
\kappa_1\circ\kappa_1\circ f_1,\dotsc,
\kappa_1\circ\kappa_n\circ f_n,
\kappa_2\circ \kappa_1\circ f_{n+1}
\colon X\to n\cdot Y+ n\cdot Y
\]
are orthogonal.
Let $\alpha\colon n\cdot Y+ Y\to (n+1)\cdot Y$ be
the associativity isomorphism.
Then
\begin{align*}
\alpha\circ(\id+\pproj_1)\circ\kappa_1\circ\kappa_i\circ f_i
&= \alpha\circ\kappa_1\circ\kappa_i\circ f_i
= \kappa_i\circ f_i \\
\alpha\circ(\id+\pproj_1)\circ\kappa_2\circ \kappa_1\circ f_{n+1}
&=\alpha\circ\kappa_2\circ\pproj_1\circ\kappa_1\circ f_{n+1}
=\kappa_{n+1}\circ f_{n+1}
\end{align*}
Therefore $(\kappa_i\circ f_i\colon X\to (n+1)\cdot Y)_{i\in\nset{n+1}}$ is orthogonal.
\end{myproof}

\begin{myproof}[Proof of Theorem~\ref{thm:chara-finpac}]
(Only if)
In a FinPAC,
the partial projections are jointly monic
by Lemma~\ref{lem:FinPAC-nary-decomp}.
To show the pullback condition,
let $f,g\colon Y\to X+X$ be arrows
with $\nabla\circ f=\pproj_1\circ g$.
Let $f_i=\pproj_i\circ f$
and $g_i=\pproj_i\circ g$ ($i=1,2$).
Using Lemma~\ref{prop:unique-pcm-enrichment},
one has $f_1\perp f_2$, $g_1\perp g_2$,
and $f_1\ovee f_2=\nabla\circ f= \pproj_1\circ g=g_1$,
so that $f_1, f_2, g_2$ are orthogonal.
By (ternary) untying,
$\kappa_1\circ f_1,\kappa_2\circ f_2,\kappa_3\circ g_2\colon
Y\to X+X+X$ are orthogonal.
Writing $\alpha\colon X+X+X\to (X+X)+X$
for the associativity isomorphism,
define $h\colon Y\to (X+X)+X$ by
\begin{align*}
h&=\alpha\circ(\kappa_1\circ f_1\ovee
\kappa_2\circ f_2\ovee
\kappa_3\circ g_2) \\
&=
\kappa_1\circ\kappa_1\circ f_1\ovee
\kappa_1\circ\kappa_2\circ f_2\ovee
\kappa_2\circ g_2 \\
&=
\kappa_1\circ f\ovee
\kappa_2\circ g_2
\qquad\text{(note $f=\kappa_1\circ f_1\ovee
\kappa_2\circ f_2$)}
\enspace.
\end{align*}
Then $\pproj_1\circ h=f$ easily,
and
\[
(\nabla+\id)\circ h=
\kappa_1\circ\nabla\circ f\ovee \kappa_2\circ g_2=
\kappa_1\circ g_1\ovee \kappa_2\circ g_2=g
\enspace.
\]
Hence $h$ is a desired mediating map.
To see the uniqueness, let $k\colon Y\to (X+X)+X$
be an arrow with $\pproj_1\circ k=f$
and $(\nabla+\id)\circ k=g$.
Then $\pproj_2\circ k=\pproj_2\circ(\nabla+\id)\circ k
=\pproj_2\circ g$.
Such $k$ is unique since the partial projections
$\pproj_1$ and $\pproj_2$ are jointly monic.

(If) Assume that a category $\catC$ satisfies the given conditions.
The joint monicity of partial projections
allows us to define the partial sum $\ovee$
on homsets $\catC(X,Y)$ in the way of
Proposition~\ref{prop:unique-pcm-enrichment}.
We show that $\catC$ is $\PCM$-enriched
with zero arrows as neutral elements.

\textbf{Associativity.}
Let $f,g,h\in\catC(X,Y)$ be arrows
with $f\perp g$ (i.e.\ compatible) via $b\colon X\to Y+Y$,
and $f\ovee g\perp h$ via $c\colon X\to Y+Y$.
By definition we have
$\nabla\circ b=f\ovee g=\pproj_1\circ c$,
so that we obtain a mediating map $d$ as in the diagram:
\[
\xymatrix@R-1pc{
X\ar@(r,u)[drr]^c
\ar@(d,l)[ddr]_b
\ar@{-->}[dr]^d
&& \\
&(Y+Y)+Y \ar[d]_{\pproj_1}
\ar[r]^-{\nabla+\id}
\pb{315} &
Y+Y \ar[d]^{\pproj_1} \\
& Y+Y\ar[r]^-{\nabla} & Y
}
\]
Then, it is straightforward to check that
\[
g\perp h
\;\;\text{via}\;\;
X\xrightarrow{d}
(Y+Y)+Y
\xrightarrow{\pproj_2+\id} Y+Y
\enspace;\enspace
\text{and}\quad
f\perp g\ovee h
\;\;\text{via}\;\;
X\xrightarrow{d}
(Y+Y)+Y
\xrightarrow{\cotup{\id,\kappa_2}} Y+Y
\enspace.
\]
Finally we have
\[
f\ovee (g\ovee h)
= \nabla\circ \cotup{\id,\kappa_2}\circ d
= \cotup{\nabla,\id}\circ d
= \nabla\circ(\nabla+\id)\circ d
= \nabla\circ c
= (f\ovee g)\ovee h
\enspace.
\]

\textbf{Commutativity.}
Let $f,g\in\catC(X,Y)$ be arrows with
$f\perp g$ via $b\colon X\to Y+Y$.
Then it is easy to see that $g\perp f$
via $\cotup{\kappa_2,\kappa_1}\circ b\colon X\to Y+Y$,
and that $g\ovee f = \nabla\circ \cotup{\kappa_2,\kappa_1}\circ b
=\nabla\circ b=f\ovee g$.

\textbf{Zero.}
For $f\in\catC(X,Y)$,
we have $0_{XY}\perp f$ via $\kappa_2\circ f\colon X\to Y+Y$,
and $0_{XY}\ovee f = \nabla\circ\kappa_2\circ f = f$.

Therefore $\catC(X,Y)$ is a PCM
for each $X,Y\in\catC$.
We need to show that the composition
$\circ\colon\catC(Y,Z)\times\catC(X,Y)\to\catC(X,Z)$
is a PCM-bihomomorphism.
Let $f\in\catC(X,Y)$
and $h,k\in\catC(Y,Z)$ be arrows with
$h\perp k$ via $b\colon Y\to Z+Z$.
Then $h\circ f\perp k\circ f$
via $b\circ f\colon X\to Z+Z$,
and $h\circ f\ovee k\circ f=\nabla\circ b\circ f = (h\ovee k)\circ f$.
We also have $0\circ f=0$. Hence $(-)\circ f$ is a PCM-homomorphism.
Next, let $h\in\catC(Y,Z)$
and $f,g\in\catC(X,Y)$ be arrows with
$f\perp g$ via $b\colon X\to Y+Y$.
Then $h\circ f\perp h\circ g$
via $(h+h)\circ b\colon X\to Z+Z$,
and $h\circ f\ovee h\circ f=\nabla\circ (h+h)\circ b=
h\circ \nabla\circ b= h\circ (f\ovee g)$.
We also have $h\circ 0=0$, and hence $h\circ(-)$
is a PCM-homomorphism.

We have shown that $\catC$ is $\PCM$-enriched.
The compatibility sum axiom holds by definition.
If $f,g\colon X\to Y$ are compatible (i.e.\ orthogonal)
via $b\colon X\to Y+Y$,
then $\kappa_1\circ f,\kappa_2\circ g\colon X\to Y+Y$
are compatible via $(\kappa_1+\kappa_2)\circ b\colon X\to (Y+Y)+(Y+Y)$.
Hence the untying axiom holds.
\end{myproof}

%===========================================================================
\section{Proof of a 2-equivalence of the 2-categories of effectuses and
FinPACs with effects}
\label{append:equiv-of-effectus-P-effectus}
%===========================================================================

Note first that, by definition,
a natural transformation $\alpha\colon F\to G$
is monoidal w.r.t.\ $(+,0)$
if the following diagrams commute.
\[
\xymatrix@R-1pc@C+1pc{
FX+FY \ar[d]_{\cotup{F\kappa_1,F\kappa_2}}
\ar[r]^{\alpha_X+\alpha_Y} &
GX+GY \ar[d]^{\cotup{G\kappa_1,G\kappa_2}} \\
F(X+Y) \ar[r]^{\alpha_{X+Y}}& G(X+Y)
}
\qquad
\xymatrix@R-1pc@C+1pc{
0 \ar[d]_{\invbang}
\ar[dr]^{\invbang} & \\
F0 \ar[r]^{\alpha_{0}}&
G0
}
\]
Obviously, the right-hand diagram always commutes.
Hence $\alpha$ is $(+,0)$-monoidal
if and only if the left-hand diagram commutes, i.e.\ when
$\alpha_{X+Y}\circ F\kappa_1=G\kappa_1\circ \alpha_X$
and $\alpha_{X+Y}\circ F\kappa_2=G\kappa_2\circ \alpha_Y$.

Let $F\colon\catA\to\catB$
be a functor between effectuses in $\Eff$,
i.e.\ a functor that preserves $1$ and $(0,+)$.
Then, the canonical arrow
$FX+1\to F(X+1)$ in the diagram below is an isomorphism.
\[
\xymatrix@R-1pc{
FX \ar[r]^-{\kappa_1} \ar[dr]_{F\kappa_1} &
FX + 1 \ar@{-->}[d] & \ar[l]_-{\kappa_2} 1 \\
& F(X+1) & \ar[l]_-{F\kappa_2} F1 \ar[u]_\cong
}
\]
We denote the inverse $F(X+1)\to FX+1$ by $l_{F,X}$
or simply by $l_X$.
We then have the following equations, which will be used repeatedly.
\begin{align}
\label{eq:lFX-kappa1}
l_{F,X}\circ F\kappa_1&= \kappa_1 \\
\label{eq:lFX-kappa2}
l_{F,X}\circ F\kappa_2&= \kappa_2\circ\bang_{F1}
\end{align}

\begin{mylemma}
\label{lem:lifting-lift-monad-preserve-functor}
Let $F\colon\catA\to\catB$ be
a functor between effectuses in $\Eff$.
Then we have a functor
$\KlL{F}\colon\KlL{\catA}\to\KlL{\catB}$
which is a `lifting' of $F$
in the sense that the following diagram commutes.
\[
\xymatrix@R-1pc@C+1pc{
\KlL{\catA} \ar[r]^{\KlL{F}} & \KlL{\catB} \\
\catA \ar[u]^{(\Lin{-})} \ar[r]^{F} & \catB \ar[u]_{(\Lin{-})}
}
\]
\end{mylemma}
\begin{myproof}
We define $\KlL{F}$ by $\KlL{F}X=FX$
and $\KlL{F}(f)=l_{F,Y} \circ Ff\colon FX\to FY+1$
for $f\colon X\Lto Y$ in $\KlL{\catA}$,
i.e.\ $f\colon X\to Y+1$ in $\catA$.
For $h\colon X\to Y$ in $\catA$,
using \eqref{eq:lFX-kappa1},
\[
\KlL{F}\Lin{h}
=l_Y\circ F\kappa_1\circ Fh
=\kappa_1\circ Fh
=\Lin{Fh}
\enspace.
\]
Therefore $\KlL{F}$ is a lifting of $F$.
Taking $h=\id_X$ we obtain $\KlL{F}(\Lid_X)=\Lid_{FX}$.
Let $f\colon X\Lto Y$ and $g\colon Y\Lto Z$
be arrows in $\KlL{\catA}$.
Note that we have
\begin{align*}
l_Z\circ F\cotup{g,\kappa_2}\circ F\kappa_1
&= l_Z\circ Fg \\
&= \cotup{l_Z\circ Fg,\kappa_2}\circ \kappa_1 \\
&= \cotup{l_Z\circ Fg,\kappa_2}\circ l_Y\circ F\kappa_1
&&\text{by~\eqref{eq:lFX-kappa1}}
\\
l_Z\circ F\cotup{g,\kappa_2}\circ F\kappa_2
&= l_Z\circ F\kappa_2 \\
&= \kappa_2\circ\bang_{F1}
&&\text{by~\eqref{eq:lFX-kappa2}} \\
&= \cotup{l_Z\circ Fg,\kappa_2}\circ \kappa_2\circ\bang_{F1} \\
&= \cotup{l_Z\circ Fg,\kappa_2}\circ l_Y\circ F\kappa_2
&&\text{by~\eqref{eq:lFX-kappa2}}
\enspace.
\end{align*}
Because $F$ preserves finite coproducts
and hence $FY\stackrel{F\kappa_1}{\longto} F(Y+1)
\stackrel{F\kappa_2}{\longgets} F1$ is a coproduct in
$\catB$, we obtain
$l_Z\circ F\cotup{g,\kappa_2}
= \cotup{l_Z\circ Fg,\kappa_2}\circ l_Y$.
Then,
\[
\KlL{F}(g\Lcirc f)
= l_Z\circ F\cotup{g,\kappa_2}\circ F f
= \cotup{l_Z\circ Fg,\kappa_2}\circ
l_Y\circ Ff
= \KlL{F}g\Lcirc \KlL{F}f
\enspace.
\]
Therefore $\KlL{F}$ is a functor.
\end{myproof}

\begin{mylemma}
The mapping $\catB\mapsto(\KlL{\catB},1)$
for an effectus $\catB$
gives rise to a 2-functor $\KlL{(-)}\colon\Eff\to\FPE$.
\end{mylemma}
\begin{myproof}
Recall $(\KlL{\catB},1)$ is a FinPAC with effects
by Theorem~\ref{thm:eff-to-Peff}.
For an arrow $F\colon \catA\to\catB$ in $\Eff$,
we have a functor $\KlL{F}\colon \KlL{\catA}\to\KlL{\catB}$
by Lemma~\ref{lem:lifting-lift-monad-preserve-functor}.
Since $\KlL{F}$ is a lifting of $F$,
the functor $\KlL{F}$ preserves finite coproducts as $F$ does.
The arrow $1_{F1}=\Lin{\bang}_{F1}\colon F1\Lto 1$
is an isomorphism because $\bang_{F1}\colon F1\to 1$ is
an isomorphism.
Since $\KlL{F}1_X=\KlL{F}\Lin{\bang}_X=\Lin{F\bang_X}$ is total,
we have $1_{F1}\Lcirc \KlL{F}1_X=1_{FX}$.
Therefore $\KlL{F}$ is an arrow in $\FPE$.

Let $F,G\colon \catA\to\catB$ be arrows
and $\alpha\colon F\To G$ a 2-cell in $\Eff$.
Note the equation
\begin{equation}
\label{eq:lFX-alpha}
(\alpha_X+\id_1)\circ l_{F,X}=l_{G,X}\circ\alpha_{X+1}
\enspace,
\end{equation}
which holds because
\begin{align*}
(\alpha_X+\id)\circ l_{F,X}\circ F\kappa_1
&= (\alpha_X+\id)\circ\kappa_1
&&\text{by \eqref{eq:lFX-kappa1}} \\
&= \kappa_1\circ\alpha_X \\
&= l_{G,X}\circ G\kappa_1\circ\alpha_X
&&\text{by \eqref{eq:lFX-kappa1}} \\
&= l_{G,X}\circ\alpha_{X+1}\circ F\kappa_1
&&\text{since $\alpha$ is $(+,0)$-monoidal}
\\
(\alpha_X+\id)\circ l_{F,X}\circ F\kappa_2
&= (\alpha_X+\id)\circ\kappa_2\circ\bang_{F1}
&&\text{by \eqref{eq:lFX-kappa2}} \\
&= \kappa_2\circ\bang_{F1} \\
&= \kappa_2\circ\bang_{G1}\circ\alpha_1 \\
&= l_{G,X}\circ G\kappa_2\circ \alpha_1
&&\text{by \eqref{eq:lFX-kappa2}} \\
&= l_{G,X}\circ\alpha_{X+1}\circ F\kappa_2
&&\text{since $\alpha$ is $(+,0)$-monoidal}
\enspace.
\end{align*}
We define $\KlL{\alpha}\colon\KlL{F}\To\KlL{G}$
by $(\KlL{\alpha})_X=\Lin{\alpha}_X\colon FX\Lto GX$.
It is natural: for $f\colon X\Lto Y$ in $\KlL{\catA}$,
\begin{align*}
(\KlL{\alpha})_Y\Lcirc\KlL{F}f
&=\Lin{\alpha}_Y\Lcirc
(l_{F,Y}\circ Ff) \\
&=(\alpha_Y+\id_1)\circ
l_{F,Y}\circ Ff \\
&=l_{G,Y}\circ\alpha_{X+1}\circ Ff
&&\text{by \eqref{eq:lFX-alpha}} \\
&=l_{G,Y}\circ G f\circ\alpha_X
&&\text{by naturality of $\alpha$} \\
&=(l_{G,Y}\circ G f)\Lcirc\Lin{\alpha}_X \\
&=\KlL{G}f\Lcirc(\KlL{\alpha})_X
\enspace.
\end{align*}
It is monoidal with respect to $(+,0)$:
\begin{align*}
(\KlL{\alpha})_{X+Y}\Lcirc \KlL{F}\Lcopr_1
&= \Lin{\alpha}_{X+Y}\Lcirc \Lin{F\kappa_1} \\
&= (\alpha_{X+Y}\circ F\kappa_1)\Linsym \\
&= (G\kappa_1\circ \alpha_X)\Linsym
&&\text{since $\alpha$ is $(+,0)$-monoidal} \\
&= \Lin{G\kappa_1}\Lcirc\Lin{\alpha}_X \\
&= \KlL{G}\Lcopr_1\Lcirc(\KlL{\alpha})_X
\end{align*}
and similarly we have
$(\KlL{\alpha})_{X+Y}\Lcirc \KlL{F}\Lcopr_2=
\KlL{G}\Lcopr_2\Lcirc(\KlL{\alpha})_Y$.
The arrow $(\KlL{\alpha})_1=\Lin{\alpha}_1\colon
F1\Lto G1$ is total,
hence $1_{G1}\Lcirc (\KlL{\alpha})_1=1_{F1}$.
Therefore $\KlL{\alpha}$ is a 2-cell in $\FPE$.

We then check that $\KlL{(-)}\colon \Eff(\catA,\catB)\to
\FPE(\KlL{\catA},\KlL{\catB})$ is a (1-)functor.
For the identity $\id_F\colon F\To F$
we have $(\KlL{(\id_F)})_X=\Lin{\id}_X\colon FX\Lto FX$,
so that $\KlL{(\id_F)}=\id_{\KlL{F}}\colon \KlL{F}\To \KlL{F}$.
Let $\alpha\colon F\To G$ and
$\beta\colon G\To H$ be 2-cells in $\Eff$.
Then
\[
(\KlL{(\beta\circ\alpha)})_X
=(\beta_X\circ\alpha_X)\Linsym
=\Lin{\beta}_X\Lcirc\Lin{\alpha}_X
=(\KlL{\beta})_X\Lcirc(\KlL{\alpha})_X
=(\KlL{\beta}\circ\KlL{\alpha})_X
\enspace.
\]
Therefore $\KlL{(\beta\circ\alpha)}=
\KlL{\beta}\circ\KlL{\alpha}$.

Now we show that $\KlL{(-)}$ is a 2-functor.
For the identity functor $\id_\catB\colon\catB\to\catB$,
it is easy to see the canonical isomorphism
$l_{\id_\catB,X}\colon X+1\to X+1$ is the identity,
so that $\KlL{(\id_\catB)}=\id_{\KlL{\catB}}$.
Let $F\colon\catA\to\catB$ and
$G\colon\catB\to\catC$ be arrows in $\Eff$.
Note the equation
\begin{equation}
\label{eq:lGFX}
l_{GF,X}=l_{G,FX}\circ G l_{F,X}
\enspace,
\end{equation}
which holds because
\begin{gather*}
l_{G,FX}\circ G l_{F,X}\circ GF\kappa_1
= l_{G,FX}\circ G \kappa_1
= \kappa_1
= l_{GF,X} \circ GF\kappa_1
\\
l_{G,FX}\circ G l_{F,X}\circ GF\kappa_2
= l_{G,FX}\circ G \kappa_2\circ G\bang_{F1}
= \kappa_2\circ\bang_{G1}\circ G\bang_{F1}
= \kappa_2\circ\bang_{GF1}
= l_{GF,X} \circ GF\kappa_2
\enspace,
\end{gather*}
using~\eqref{eq:lFX-kappa1} and~\eqref{eq:lFX-kappa2}.
For $f\colon X\Lto Y$ in $\KlL{\catA}$,
using \eqref{eq:lGFX},
\[
\KlL{(GF)}f
=l_{GF,Y} \circ GF f
=l_{G,FY}\circ Gl_{F,Y} \circ GF f
=\KlL{G}(l_{F,Y} \circ F f)
=\KlL{G}\KlL{F}f
\enspace.
\]
Hence $\KlL{(GF)}=\KlL{G}\KlL{F}$.
For $\alpha\colon F\To F'$, one has
\[
(\KlL{(G\alpha)})_X
=\Lin{G\alpha_X}
=\KlL{G}\Lin{\alpha}_X
=(\KlL{G}\KlL{\alpha})_X
\enspace,
\]
so that $\KlL{(G\alpha)}=\KlL{G}\KlL{\alpha}$.
For $\alpha\colon G\To G'$, one has
\[
(\KlL{(\beta F)})_X
=\Lin{\beta}_{FX}
=(\KlL{\beta})_{\KlL{F}X}
=(\KlL{\beta}\KlL{F})_X
\enspace,
\]
and therefore $\KlL{(\beta F)}=\KlL{\beta}\KlL{F}$.
\end{myproof}

\begin{mylemma}
The mapping $(\catC,I)\mapsto\totcat{\catC}$
for a FinPAC with effects $(\catC,I)$
gives rise to a 2-functor\linebreak $\totcat{(-)}\colon\FPE\to\Eff$.
\end{mylemma}
\begin{myproof}
Recall that $\totcat{\catC}$ is an effectus
by Theorem~\ref{thm:Peff-to-eff}.
Let $F\colon(\catC,I_\catC)\to(\catD,I_\catD)$
be an arrow in $\FPE$.
If $f\colon X\to Y$ is a total arrow in $\catC$,
then
\[
\Dp(Ff)
=1_{FY}\circ Ff
=1_{FI}\circ F 1_Y\circ Ff
=1_{FI}\circ F \Dp(f)
=1_{FI}\circ F 1_X
=1_{FX}
\enspace,
\]
that is, $Ff$ is total.
Therefore $F$ restricts to the functor
$\totcat{F}\colon\totcat{\catC}\to\totcat{\catD}$
in a commutative diagram:
\[
\xymatrix@R=.8pc@C=3pc{
\totcat{\catC}\ar@{^(->}[d]
\ar[r]^{\totcat{F}} &
\totcat{\catD} \ar@{^(->}[d] \\
\catC \ar[r]^F & \catD
}
\]
Because $\totcat{\catC}$ inherits coproducts from
$\catC$, the functor $\totcat{F}$ preserves finite coproducts
as $F$ does.
Recall that $I_\catC$ and $I_\catD$ are the final objects in
$\totcat{\catC}$ and $\totcat{\catD}$ respectively.
By definition we have $FI_\catC\cong I_\catD$
in $\catD$ and hence in $\totcat{\catD}$,
so that $\totcat{F}$ preserves the final object.
Therefore $\totcat{F}$ is an arrow in $\Eff$.

Let $F,G\colon\catC\to\catD$ be arrows
and $\alpha\colon F\To G$ a 2-cell in $\FPE$.
Then
\begin{align*}
\Dp(\alpha_X)
= 1_{GX}\circ\alpha_X
&= 1_{GI_\catC}\circ G1_X\circ\alpha_X
&&\text{since $G$ is an arrow in $\FPE$} \\
&= 1_{GI_\catC}\circ \alpha_{I_\catC} \circ F 1_X
&&\text{by naturality of $\alpha$}\\
&= 1_{FI_\catC}\circ F 1_X
&&\text{since $\alpha$ is a 2-cell in $\FPE$} \\
&= 1_{FX}
&&\text{since $F$ is an arrow in $\FPE$}
\enspace,
\end{align*}
so that $\alpha_X$ is total.
Hence we can restrict $\alpha$
to the natural transformation
$\totcat{\alpha}\colon\totcat{F}\To\totcat{G}$
with $(\totcat{\alpha})_X=\alpha_X$,
which is obviously a 2-cell in $\Eff$.
Then it is easy to see that $\totcat{(-)}$
gives a (1-)functor $\FPE(\catC,\catD)\to\Eff(\totcat{\catC},\totcat{\catD})$.

Finally, we can easily check
$\totcat{(\id_\catC)}=\id_{\totcat{\catC}}$,
$\totcat{(GF)}=\totcat{G}\totcat{F}$,
$\totcat{(G\alpha)}=\totcat{G}\totcat{\alpha}$,
$\totcat{(\beta F)}=\totcat{\beta}\totcat{F}$
for arrows $F\colon\catC\to\catD$
and $G\colon\catD\to\catE$,
and 2-cells $\alpha\colon F\To F'$ and $\beta\colon G\To G'$
in $\FPE$.
Therefore $\totcat{(-)}$ gives a 2-functor $\FPE\to\Eff$.
\end{myproof}

\begin{mytheorem}
The 2-functors $\KlL{(-)}\colon\Eff\to\FPE$
and $\totcat{(-)}\colon\FPE\to\Eff$
form a 2-equivalence of 2-categories $\Eff\simeq\FPE$.
Namely, there are 2-natural isomorphisms
$\id_\Eff\cong \totcat{(\KlL{(-)})}$
and $\id_\FPE\cong \KlL{(\totcat{(-)})}$.
\end{mytheorem}
\begin{myproof}
We write $\Phi_\catB\colon\catB\to\totcat{(\KlL{\catB})}$
for the isomorphism of categories
in Proposition~\ref{prop:isom-B-Bot},
which is given by $\Phi_\catB X=X$ and
$\Phi_\catB f=\Lin{f}$.
It preserves finite coproducts and the
final object, so that $\Phi_\catB$ is an arrow in $\Eff$.
Let $F\colon\catA\to\catB$ be an arrow in $\Eff$.
Because $\KlL{F}$ is a lifting of $F$,
and $\totcat{(\KlL{F})}$ is a restriction of $\KlL{F}$,
the following diagram commutes.
\[
\xymatrix@R-1pc@C+1pc{
\catA\ar[r]^F \ar[d]^\cong_{\Phi_\catA} &
\catB\ar[d]_\cong^{\Phi_\catB} \\
\totcat{(\KlL{\catA})}
\ar[r]^{\totcat{(\KlL{F})}} &
\totcat{(\KlL{\catB})}
}
\]
Let $\alpha\colon F\To G$ be a 2-cell in $\Eff$.
Then
\[
(\totcat{(\KlL{\alpha})}\Phi_\catA)_X
=(\totcat{(\KlL{\alpha})})_{\Phi_\catA X}
=(\KlL{\alpha})_X
=\Lin{\alpha}_X
=\Phi_\catB\alpha_X
=(\Phi_\catB\alpha)_X
\enspace,
\]
so that $\totcat{(\KlL{\alpha})}\Phi_\catA=\Phi_\catB\alpha$.
Therefore $\Phi$ defines a 2-natural isomorphism $\id_\Eff\To\totcat{(\KlL{(-)})}$.

Next, we write $\Psi_\catC\colon\KlL{(\totcat{\catC})}\to\catC$
for the isomorphism of categories in Lemma~\ref{lem:P-effectus-coproduct},
which is defined by $\Psi_\catC X=X$ and $\Psi_\catC f=\pproj_1\circ f$.
It preserves finite coproducts and the unit object $I$,
since $I$ is the final object of $\totcat{\catC}$.
If we write $1_X$ for the top of $\catC(X,I)$,
then the top of $\KlL{(\totcat{\catC})}(X,I)$
is $\Lin{1}_X=\kappa_1\circ 1_X$,
and therefore we have $\Psi_\catC \Lin{1}_X=\pproj_1\circ\kappa_1\circ 1_X=1_X$.
Hence $\Psi_\catC$ is an arrow in $\FPE$.
Let $F\colon\catC\to\catD$ be an arrow in $\FPE$.
Note that the following diagram commutes,
\[
\xymatrix@R-1pc{
F(X+I_\catC)\ar[r]^-{l_{\totcat{F},X}}
\ar[dr]_{F\pproj_1} &
FX+I_\catD\ar[d]^{\pproj_1} \\
& FX
}
\]
since
\begin{gather*}
\pproj_1\circ l_{\totcat{F},X} \circ F\kappa_1
=\pproj_1\circ \kappa_1
=\id_{FX}
=F\id_X
=F\pproj_1\circ F\kappa_1
\\
\pproj_1\circ l_{\totcat{F},X} \circ F\kappa_2
=\pproj_1\circ \kappa_2\circ 1_{FI_\catC}
=0_{FI_\catC,FX}
=F0_{I_\catC,X}
=F\pproj_1\circ F\kappa_2
\enspace,
\end{gather*}
where $F0_{I_\catC,X}=0_{FI_\catC,FX}$ holds
because $F$ preserves the zero object.
For $f\colon X\Lto Y$ in $\KlL{(\totcat{\catC})}$,
we have
\[
\Psi_\catD\KlL{(\totcat{F})} f
=\pproj_1\circ l_{\totcat{F},Y}\circ Ff
=F\pproj_1\circ Ff
=F(\pproj_1\circ f)
=F\Psi_\catC f
\enspace,
\]
and hence $\Psi_\catD\KlL{(\totcat{F})}=F\Psi_\catC$.
Let $\alpha\colon F\To G$ be a 2-cell in $\FPE$.
Then
\[
(\Psi_\catD\KlL{(\totcat{\alpha})})_X
=\Psi_\catD(\KlL{(\totcat{\alpha})})_X
=\pproj_1\circ \Lin{\alpha}_X
=\alpha_X
=\alpha_{\Psi_\catC X}
=(\alpha\Psi_\catC)_X
\enspace,
\]
so that $\Psi_\catD\KlL{(\totcat{\alpha})}=\alpha\Psi_\catC$.
Therefore $\Psi$ defines a 2-natural isomorphism
$\KlL{(\totcat{(-)})}\To\id_\FPE$.
\end{myproof}

%==================================
\section{Convex sets over a division effect monoid}
\label{append:convex-set}
%==================================

Throughout this section,
we let $M$ be a division effect monoid
(see Definition~\ref{def:division-emon}).

\begin{mylemma}
\label{lem:div-emon-cancel}
For $r,s,t,u\in M$ with $r\le s$, $s\cdot t\le u$,
$s\ne0$ and $u\ne0$, one has
$(r/s)\cdot(st/u)=rt/u$.
In particular $(r/s)\cdot(s/u)=r/u$,
by setting $t=1$.
\end{mylemma}
\begin{myproof}
Since $(r/s)\cdot(st/u)\cdot u
=(r/s)\cdot s\cdot t
=r\cdot t$.
\end{myproof}

\begin{mylemma}
\label{lem:div-emon-hom}
For each nonzero $t\in M$,
the `multiplication by $t$' map
$(-)\cdot t\colon M\to{\downarrow}(t)$
is an effect module (over $M$) isomorphism,
with the inverse $(-)/t\colon {\downarrow}(t)\to M$.
In particular, $(-)/t$ is an effect module homomorphism:
$0/t=0$; $t/t=1$; $(r\ovee s)/t=r/t\ovee s/t$;
and $(rs/t)=r(s/t)$.
\end{mylemma}
\begin{myproof}
The definition of division says that
the map $(-)\cdot t\colon M\to{\downarrow}(t)$ is bijective.
It is easy to see that $(-)\cdot t$ is an effect module homomorphism.
Therefore, to prove it is an isomorphism,
it suffices to show that it reflects the orthogonality:
if $r\cdot t\perp s\cdot t$ and
$r\cdot t\ovee s\cdot t\le t$, then $r\perp s$.
Since the case $r=0$ is trivial,
we assume $r\ne 0$. Then $r\cdot t$ is nonzero too
because $(-)\cdot t\colon M\to{\downarrow}(t)$ is bijective.
Note that $s^\bot\cdot t=t\ominus s\cdot t\ge r\cdot t$
and hence $s^\bot\cdot t$ is nonzero as well.
Then
\[
r=(r\cdot t)/t
=((r\cdot t)/(s^\bot\cdot t)) \cdot ((s^\bot\cdot t)/t)
=((r\cdot t)/(s^\bot\cdot t)) \cdot s^\bot
\le s^\bot
\enspace,
\]
so that $r\perp s$.
\end{myproof}

The division allows us to construct coproducts in the category $\Conv_M$ explicitly,
in the same way as the case $M=[0,1]$ done in \cite{JacobsWW2015}.
First we construct a coproduct of the form $X+1$.
For a convex set $X$ over $M$,
we define a ``lifted'' convex set $X_\bullet$ as follows.
\[
X_\bullet
=\{(x,r)\in (X\cup\{\bullet\})\times M\mid
x=\bullet \iff r=0\}
\;\;(=X\times (M\setminus\{0\}) \cup \{({\bullet},0)\})
\]
For a formal convex sum $\sum_i\ket{(x_i,r_i)}s_i\in\Dst_M (X_\bullet)$,
define the actual sum by
\begin{equation}
\label{eq:def-lifted-convex-sum}
\textstyle
\bigovee_i (x_i,r_i)s_i
= \bigl(\bigovee_i x_i(r_is_i/ t),t\bigr)
\qquad\text{where:}\enspace
t=\bigovee_i r_is_i
\enspace.
\end{equation}
Note that $\bigovee_i(r_is_i/t)=(\bigovee_i r_is_i)/t=t/t=1$.
The formula~\eqref{eq:def-lifted-convex-sum}
is not completely rigorous in the case $t=0$ or $x_i=\bullet$,
but the meaning will be clear.
For example, we often mean $(\bullet,0)$ by
$(e,0)$ even when $e$ is an expression that does not make sense.
Then, the diagram
\[
\xymatrix{
X\ar[r]^{\kappa_1} & X_\bullet &
\ar[l]_{\kappa_2} 1
}
\qquad\text{where:}\enspace
\kappa_1(x)=(x,1)
\; ;\enspace
\kappa_2(\bullet)=(\bullet,0)
\]
is a coproduct in $\Conv_M$,
i.e.\ $X_\bullet\cong X+1$.
For $f\colon X\to Y$ and
$g\colon 1\to Y$,
define $[f,g]\colon X_\bullet\to Y$
by $[f,g](x,r)=f(x)r\ovee g(\bullet)r^\bot$.

Let $X$ and $Y$ be convex sets over $M$.
Define a (convex) subset $X+Y\subseteq X_\bullet\times Y_\bullet$
by: $((x,r),(y,s))\in X+Y$ $\Iff$ $r\perp s$ and $r\ovee s=1$.
Then the diagram
\[
\xymatrix{
X\ar[r]^-{\kappa_1} & X+Y &
\ar[l]_-{\kappa_2} Y
}
\qquad\text{where:}\enspace
\begin{aligned}
\kappa_1(x)&=((x,1),(\bullet,0)) \\
\kappa_2(y)&=((\bullet,0),(y,1))
\end{aligned}
\]
is a coproduct in $\Conv_M$.
For $f\colon X\to Z$ and $g\colon Y\to Z$,
define $[f,g]\colon X+Y\to Z$
by $[f,g]((x,r),(y,s))=f(x)r\ovee g(y)s$.

Finally, we show that
$\Conv_M$ is an effectus.
Note that we use a weaker joint monicity
requirement than~\cite[Definition~12]{JacobsWW2015}.

\begin{myproposition}
\label{prop:conv-effectus}
The category $\Conv_M$ of convex sets
over a division effect monoid $M$ is an effectus.
\end{myproposition}
\begin{myproof}
The category $\Conv_M$ has
binary coproducts as we described above.
It also has the empty convex set $0=\emptyset$ as
an initial object (unless $M$ is trivial, i.e.\ a singleton $\{0\}$;
in that case, $\Conv_M$ is a trivial category,
which is trivially an effectus),
and the singleton convex set $1$
as a final object.
The pullback requirements are shown
in the same way as the case $M=[0,1]$,
see~\cite[Proposition~15]{JacobsWW2015}.

We now prove that the maps
$\cotup{\kappa_1,\kappa_2,\kappa_2},\cotup{\kappa_2,\kappa_1,\kappa_2}\colon
1+1+1\to 1+1$ are jointly monic.
It is not hard to see that
\[
1+1\cong\{(r,s)\in M\times M\mid
r\ovee s=1\}\cong M
\]
with coprojections $\kappa_i\colon 1\to M$ given by $\kappa_1(\bullet)=1$
and $\kappa_2(\bullet)=0$. Similarly we have
\[
1+1+1\cong\{(r,s,t)\in M\times M\times M\mid
r\ovee s\ovee t=1\}
\eqqcolon M_3
\enspace.
\]
Then the maps $\cotup{\kappa_1,\kappa_2,\kappa_2},\cotup{\kappa_2,\kappa_1,\kappa_2}\colon
M_3\to M$ are given by
\[
\cotup{\kappa_1,\kappa_2,\kappa_2}(r,s,t)
=\kappa_1(\bullet)r\ovee\kappa_2(\bullet)s\ovee
\kappa_2(\bullet)t=r\ovee 0\ovee 0= r
\]
and similarly $\cotup{\kappa_2,\kappa_1,\kappa_2}(r,s,t)=s$.
To see they are jointly injective, assume
\[
\cotup{\kappa_1,\kappa_2,\kappa_2}(r,s,t)=\cotup{\kappa_1,\kappa_2,\kappa_2}(r',s',t')
\quad\text{and}\quad
\cotup{\kappa_2,\kappa_1,\kappa_2}(r,s,t)=\cotup{\kappa_2,\kappa_1,\kappa_2}(r',s',t')
\enspace.
\]
Then $r=r'$, $s=s'$ and so $t=(r\ovee s)^\bot=(r'\ovee s')^\bot=t'$.
Hence $(r,s,t)=(r',s',t')$.
\end{myproof}

\end{document}